\documentclass{aa}
\usepackage{graphics}
\begin{document}
\def\Ha{H$_{\alpha}$}
   \title{A search for late-type supergiants in the inner
regions of the Milky Way
\thanks{Based on observations collected at the Centro Astron\'omico
Hispano-Alem\'an, Calar Alto, Spain; and at the European Southern
Observatory, La Silla, Chile (programme 71.B-0274(A))}}

   \author{F. Comer\'on\inst{1}
        \and
       J. Torra\inst{2}
        \and
       C. Chiappini\inst{3}
        \and
       F. Figueras\inst{2}
        \and
       V.D. Ivanov\inst{4}
        \and
       S.J. Ribas\inst{2}
          }

   \offprints{F. Comer\'on}

   \institute{European Southern Observatory, Karl-Schwarzschild-Strasse 2,
              D-85748 Garching, Germany\\
          email: fcomeron@eso.org
          \and
          Departament d'Astronomia i Meteorologia, Universitat de
          Barcelona, Av. Diagonal 647, E-08028 Barcelona, Spain\\
          email: jordi@am.ub.es
          \and
          Osservatorio Astronomico di Trieste, Via G.B. Tiepolo 11,
          I-34131 Trieste, Italy\\
          email: chiappin@ts.astro.it
          \and
          European Southern Observatory, Alonso de Cordova, 3107,
          Santiago 19, Chile\\
          email: vivanov@eso.org
             }
   \date{Received; accepted}

\abstract{We present the results of a narrow-band infrared imaging
survey of a narrow strip ($12'$ wide) around the galactic equator
between $6^\circ$ and $21^\circ$ of galactic longitude aimed at
detecting field stars with strong CO absorption, mainly late-type
giants and supergiants. Our observations include follow-up low
resolution spectroscopy ($R = 980$) of 191 selected candidates in
the $H$ and $K$ bands. Most of these objects have photometric and
spectroscopic characteristics consistent with them being red
giants, and some display broad, strong absorption wings due to
water vapor absorption between the $H$ and $K$ bands. We also
identify in our sample 18 good supergiant candidates characterized
by their lack of noticeable water absorption, strong CO bands in
the $H$ and $K$ windows, and $HK_S$ photometry suggestive of high
intrinsic luminosity and extinctions reaching up to $A_V \simeq
40$~mag. Another 9 additional candidates share the same features
except for weak H$_2$O absorption, which is also observed among
some M supergiants in the solar neighbourhood. Interesting
differences are noticed when comparing our stars to a local sample
of late-type giants and supergiants, as well as to a sample of red
giants in globular clusters of moderately subsolar metallicity and
to a sample of bulge stars. A large fraction of the stars in our
sample have CaI and NaI features markedly stronger than those
typical in the local reference sample (both giants and
supergiants), whereas the equivalent widths of the CO bands are
similar or weaker. In this regard, our stars in the inner Milky
Way disk display differences very similar to those identified by
other authors between cool giants and supergiants near the
galactic center and their counterparts in the solar neighbourhood.
We propose that the systematic spectroscopic differences of our
inner Galaxy stars are due to their higher metallicities that
cause deeper mixing in their mantles, resulting in lower surface
abundances of C and O and higher abundances of CN, which
contribute to the strength of the CaI and NaI features at low
resolution. Our results stress the limitations of using local
stars as templates for the study of composite cool stellar
populations such as central starbursts in galaxies.
  }

   \maketitle

%

\section{Introduction}

  As compared to the large amount of information available nowadays
on the inner regions of nearby galaxies, the Population~I
component of the central kiloparsecs of the Milky Way is still
relatively unknown. Visual extinctions towards objects located in
the galactic plane only a few kiloparsecs from the galactic
center, typically reaching several tens of magnitudes, restrict
the observations to the infrared and radio domains, or to X or
gamma rays for objects associated with high energy phenomena.
Moreover, a variety of factors such as distance, extinction, field
crowding, line-of-sight confusion, or absence of distinctive
properties in the accessible wavelength range contribute to make
difficult the identification or the detailed observation of
individual objects. Observational programs such as large area
surveys probing radio continuum and recombination line emission
have supplied much of our present knowledge on the gas
photoionized by massive stars (Lockman~\cite{lockman89}, Helfand
et al.~\cite{helfand92}, Whiteoak~\cite{whiteoak92}, Becker et
al.~\cite{becker94}, Kuchar \& Bania~\cite{kuchar94},
McClure-Griffiths~\cite{mcclure01}, Kolpak et
al.~\cite{kolpak03}), and on the nonthermal emission due to recent
supernovae in the inner Galaxy (Leahy \& Wu~\cite{leahy89},
Green~\cite{green91}, \cite{green01}). Molecular line emission
surveys have provided an overall picture of the distribution and
physical conditions of giant molecular clouds (Dame et
al.~\cite{dame87}, \cite{dame01}, Chiar et al.~\cite{chiar94},
Bronfman et al.~\cite{bronfman96}, \cite{bronfman00}, McQuinn et
al.~\cite{mcquinn02}), while mid-infrared surveys such as that
carried out by IRAS have revealed a wealth of star forming sites
(Wood \& Churchwell~\cite{wood89}, Hughes \&
MacLeod~\cite{hughes94}, Codella et al.~\cite{codella95},
Comer\'on \& Torra~\cite{comeron96}, Egan et al.~\cite{egan98}).
Recent near-infrared surveys like DENIS (Ruphy et
al.~\cite{ruphy97}, L\'opez-Corredoira et al.~\cite{lopez01}, Van
Loon et al.~\cite{vanloon03}), 2MASS (Alard~\cite{alard01},
Ojha~\cite{ojha01}, L\'opez-Corredoira et al.~\cite{lopez02}) or
TMGS (L\'opez-Corredoira et al.~\cite{lopez99}, \cite{lopez01},
Picaud et al.~\cite{picaud03}) are yielding extensive lists of
stellar objects suitable for studies of galactic structure.
Valuable as all those observational approaches are, each one has
its own limitations and needs to be complemented by the others in
order to provide a global picture of the young component of the
inner parts of our Galaxy. On the other hand, this young component
is an excellent tracer of processes of primary importance
concerning the structure and the composition of our Galaxy, such
as the dynamical response of the disk to the central bar, the
location of the spiral arms, the current star forming activity, or
the radial dependency of the chemical composition, thus justifying
the efforts leading to the characterization of its properties.

  In this paper we present a study on a particularly promising,
yet relatively unexploited, Population I tracer in the inner
Galaxy. Late-type supergiant stars have interesting properties
that make them particularly suitable for the study of aspects of
the extreme Population~I component that are either inaccessible or
very elusive in other objects of similar age, such as the
structure, kinematics, and chemistry of the young galactic disk,
while their numerical abundance can provide a measurement of its
present-day massive star formation rate. They have extremely
bright magnitudes at infrared wavelengths, where the dust in the
galactic disk is far less opaque than in the visible. As it turns
out, a nearly {\it complete} census of late-type supergiants in
our Galaxy is in principle a feasible task for a moderately large
telescope equipped with a near-infrared camera, even if some of
the targets may be as far as 25~kpc away and obscured by as much
as 50~mag at $V$. While the bright magnitudes of supergiants
facilitate detection, the infrared spectral properties of cool
photospheres facilitate their identification in conveniently
designed infrared surveys. The $K$-band spectra of objects with
temperatures below $\sim 3500$~K are characterized by prominent CO
bands appearing longwards of 2.29~$\mu$m, thus allowing the
selection of candidate late-type stars (red giant branch (RGB)
stars, long period variables in the assymptotic giant branch
(AGB), and red supergiants) by comparing narrow band images
centered respectively on and off the CO band region. At low- and
medium- resolution spectroscopy, many narrow atomic features in
the $H$ and $K$ bands allow further investigation of their
intrinsic properties via their different sensitivities to
temperature, surface gravity, metallicity, and surface abundances.
Moreover, infrared observations of cool, luminous stars in the
inner galactic disk provide actual, individual stellar templates
suitable for the interpretation of the integrated spectrum of
systems undergoing massive star formation, such as central
starbursts and the environments of AGNs (e.g. Origlia et
al.~\cite{origlia93}, Kotilainen et al.~\cite{kotilainen96},
Origlia \& Oliva~\cite{origlia00}, F\"orster-Schreiber et
al.~\cite{forster01}, Alonso-Herrero et al.~\cite{alonso01}), and
are an important component in the modeling of the integrated light
of entire systems (e.g. Fioc \& Rocca-Volmerange \cite{fioc97},
Bruzual \& Charlot \cite{bruzual03}).

  This paper describes a search for late-type luminous, obscured
cool field stars in a $15^\circ$-long strip of the galactic plane
of approximately 12' width in galactic latitude, consisting of
narrow-band imaging and low resolution follow-up spectroscopy of
the best candidates identified. Although our focus is on the
identification criteria of cool supergiants we also have obtained
as a byproduct a large number of spectra of cool RGB and AGB stars
in the same general direction, which can be a useful resource for
studies of metallicity patterns across the inner galactic disk and
for the construction of templates in integrated stellar
populations spectra.

  The observations and data reduction process are described in
Section~\ref{observations}. The results are presented in
Section~\ref{results}, where we discuss the properties of our
sample and its comparison to those of reference samples containing
objects in the solar neighbourhood, in the galactic bulge, and in
relatively metal-rich globular clusters. We discuss criteria that
allow the definition of a sample of red supergiant candidates and
present a list of the best ones identified among our observed
objects. Section~\ref{discussion} discusses one of the main
results of our work, concerning systematic spectral differences
that we find between our objects and those in the two reference
samples. Section~\ref{conclusions} summarizes our results.

\section{Observations\label{observations}}

\subsection{Near-infrared imaging}

\subsubsection{Observing strategy and filter choice}

  The basis for our preliminary identification of possible cool
supergiants consists of multiband images of narrow strips parallel
to the galactic plane obtained with MAGIC, the near-infrared
camera and spectrograph at the 1.23 m telescope at the
German-Spanish Astronomical Center in Calar Alto. In imaging mode,
MAGIC yields a pixel scale of 1"15 per pixel resulting in a field
of view of $4'9 \times 4'9$ over the $256 \times 256$ pixel$^2$
array. Our imaging observations were obtained on two separate
observing runs, carried out respectively on 3-13 July 1998 and
19-21 June 2001.

  The observations consisted of obtaining sequences of short
exposures at positions separated by 1' intervals along the
galactic plane. Since the edges of the detector were aligned along
the North-South and East-West directions and the series of images
were obtained by moving the telescope along the galactic equator,
stars lying at galactic latitudes near the edge of the coverage of
the sequences were observed in fewer telescope pointings. At the
beginning of each sequence the filter and the detector integration
parameters were set, and then the sequence was started to obtain a
total of 180 overlapping fields covering an arc of $3^\circ$ in
galactic longitude. The starting positions corresponded to
galactic longitudes of $l=6^\circ$, $9^\circ$, $12^\circ$,
$15^\circ$, and $18^\circ$, and galactic latitudes of $b=0'$,
$+4'$, and $-4'$, thus allowing in principle overlap between
contiguous sequences either in galactic longitude or latitude.
Unfortunately, a major mechanical intervention on the mount of the
1.23m telescope between our two runs produced a slight
misalignment of its polar axis and an inaccurate pointing model in
our run of 2001. As a result, the sequences starting at $b =-4'$
and the one at $l = 15^\circ$, $b = +4'$ are not exactly parallel
to the galactic equator and do not overlap over their entire
lengths with contiguous sequences starting at the same galactic
longitude.

  Four filters were used: a CO filter centered at 2.295~$\mu$m
with a relative width $\Delta \lambda / \lambda = 1\%$, a
$K$-continuum narrow-band filter centered at 2.260~$\mu$m with
$\Delta \lambda / \lambda = 2.6\%$, and the broad-band filters $J$
(1.25~$\mu$m, $\Delta \lambda / \lambda = 24\%$) and $H$
(1.65~$\mu$m, $\Delta \lambda / \lambda = 18\%$). The choice of
narrow-band filters aims at allowing the detection of stars with
large flux drops due to CO absorption, which is a characteristic
feature of late-type photospheres, by comparing the images taken
through the narrow-band CO and $K$-continuum filters (see below),
while the broad-band filters were initially intended to locate
objects in color-color and color-magnitude diagrams. Exposure
times per telescope position were 9 seconds in the CO filter; 6 in
the $K$-continuum filter; 4 in the $H$ filter; and 8 in the $J$
filter. These exposure times were obtained by stacking together as
many individual exposures of 1 second each as needed. The exposure
time per sky position is variable, as it depends on the number of
telescope pointings containing that position as described above.
It ranges from one to six times the exposure time per field, the
latter value corresponding to stars lying close to the central
galactic latitude of the sequence. Each frame was combined with
the overlapping portion of the images taken before and after in
the sequence in order to filter out remaining unmasked bad pixels
or cosmic ray hits.

  In general preference was given to completing one sequence in all
filters before proceeding to the next one, in order to have
observations as closely spaced as possible. However this had to be
often overridden due to visibility constraints as the regions
imaged, lying at declinations between $-24^\circ$ and $-10^\circ$,
are visible few hours per night from a mid-Northern site like
Calar Alto. A consequence of this is that observations through the
different filters are not consecutive in some cases, but separated
by an interval of two days at most. Such time differences are
nevertheless negligible by comparison to typical variability
periods of evolved red stars, and we thus considered them entirely
acceptable for the goals of our project.

\subsubsection{Data reduction and calibration}

  The reduction of our imaging data, as well as of our spectra as
described in Section~\ref{obs_spectra}, was carried out using
IRAF\footnote{IRAF is distributed by NOAO, which is operated by
the Association of Universities for Research in Astronomy, Inc.,
under contract to the National Science Foundation} tasks and
dedicated scripts. The images were reduced using dome flat fields
in each filter that were used also to construct bad pixel masks.
Dark-subtracted, flat-fielded and masked images were
sky-subtracted by constructing a sky frame made out of the 20
frames taken closest in time, conveniently median-averaged and
with the stellar images removed by sigma clipping. Next, the
images obtained within each of the $b = 0^\circ$ sequences in the
CO on-band filter, which we took as a reference, were mosaicked
into a single large image. To this end, accurate offsets between
consecutive pointings in a sequence were determined by
automatically finding stars in the common area of overlapping
images and using them as a reference. In case that a contiguous
sequence (i.e., one starting at the same galactic longitude but a
different latitude) was obtained during the same run, a similar
procedure was used to align its component images with the
reference $b = 0^\circ$ sequence: the position of each individual
image along the parallel sequence with respect to the reference
one was determined by using stars in the overlapping areas. This
was not possible when sequences starting at the same galactic
longitude were obtained on different runs, due to the pointing
model inaccuracy outlined above. In such cases the sequence
obtained on the second run was combined separately. The combined
image in the CO filter was used as the reference for the alignment
of each frame in the other filters, thus avoiding large
misalignments between images in different filters due to
accumulated small errors in the registering of consecutive images
in each of them.

  Due to the crowdedness of the fields in the galactic plane and
the undersampling of the point-spread function, digital photometry
with an undersized aperture of 2 pixel (2"3) radius was found to
provide the best results. To compensate for the variability of the
sky transparency conditions and for the variable image quality
during the execution of any given sequence, instrumental zeropoint
variations along the sequence were determined by comparing the
digital photometry on the brightest nonsaturated stars in the area
common to each pair of overlapping frames. Zeropoints for each
sequence in the $J$, $H$, and $K$-continuum bands were estimated
by comparing our instrumental photometry of bright nonsaturated
stars to the 2MASS catalog, assuming that the narrow-band
$K$-continuum magnitude can be approximated by the broad-band
2MASS $K_S$ magnitude.

  The outlined procedure for photometric calibration is appropriate
to yield moderate accuracies in the $J$, $H$, and $K$-continuum
bands that are sufficient for our purposes: a comparison of the
instrumental and 2MASS magnitudes show typical scatters of $\sim
0.1$~mag around the average zeropoints, and a drift of the
zeropoint of the same order at most along a given sequence.
However, scatters of that order are too large for an appropriate
measurement of the flux drop due to CO absorption, that we use as
the primary criterion to select candidate cool stars in our
sample. We thus used instead a self-calibrated method based on the
fact that, in any random stellar field, most of the stars are
expected to have a negligible drop in the flux at the position of
the CO bands. We calculated local zeropoints of the photometric CO
index (defined as ${\rm [CO]_{phot}} = m(2.295) - m(2.26)$, where
$m(2.295)$ and $m(2.26)$ are the magnitudes in the CO and
$K$-continuum filters respectively) at each longitude by assuming
that its value averaged over a large number of randomly selected
stars must be very close to zero. This average was calculated
considering the stars located within $|\Delta l| = 10'$ of each
longitude point and on the same galactic longitude band, and with
weights inversely proportional to the square of the probable error
in magnitude of each contributing star. The calculation was done
in two steps, the first one including all the stars in the average
and the second removing stars with a highly deviant ${\rm
[CO]_{phot}}$. In this way, cool stars that could skew the
distribution of ${\rm [CO]_{phot}}$ values could be selected as
those stars standing out due to a value of that index greater than
a given threshold above the average value of the surrounding
stars.

  A master list of detected stars was obtained by coadding the
images taken in all the filters and running DAOFIND (Stetson
\cite{stetson87}) on the combined image. With this master list of
stellar positions, we performed digital photometry with DAOPHOT on
the images in each filter to obtain the multiband photometry of
each star detected in the combined image. Many stars were not
detected in some bands, either because they fell just outside the
edges of the image due to small differences between the starting
position of the sequence in each filter, because they were too
faint in that filter to be detected, or because of confusion with
a nearby source in the case of the images obtained under poorer
seeing conditions. Finally, approximate stellar positions in J2000
coordinates were determined by using as a reference the positions
of unobscured stars in our images that could be identified in
Digitized Sky Survey, and whose equatorial coordinates were
available from the USNO astrometric catalog.

  We must remark at this point that the conditions under which our
observations were made resulted in an uneven depth of the catalog
and other factors affecting its completeness. Seeing conditions
during the execution of the observations ranged from $\sim 1''$ to
over $3''$, thus resulting in a regional dependence of the source
confusion limits. Likewise, part of our sequences was affected by
the presence of high altitude dust combined with the need for
observing often at high airmasses, producing regional dependencies
of the depth of the frames. The net effects of all these
circumstances are rather uncertain and difficult to quantify, a
fundamental limitation of the present survey that must be kept in
mind.

\begin{figure}
   \resizebox{8.5cm}{!}{\includegraphics{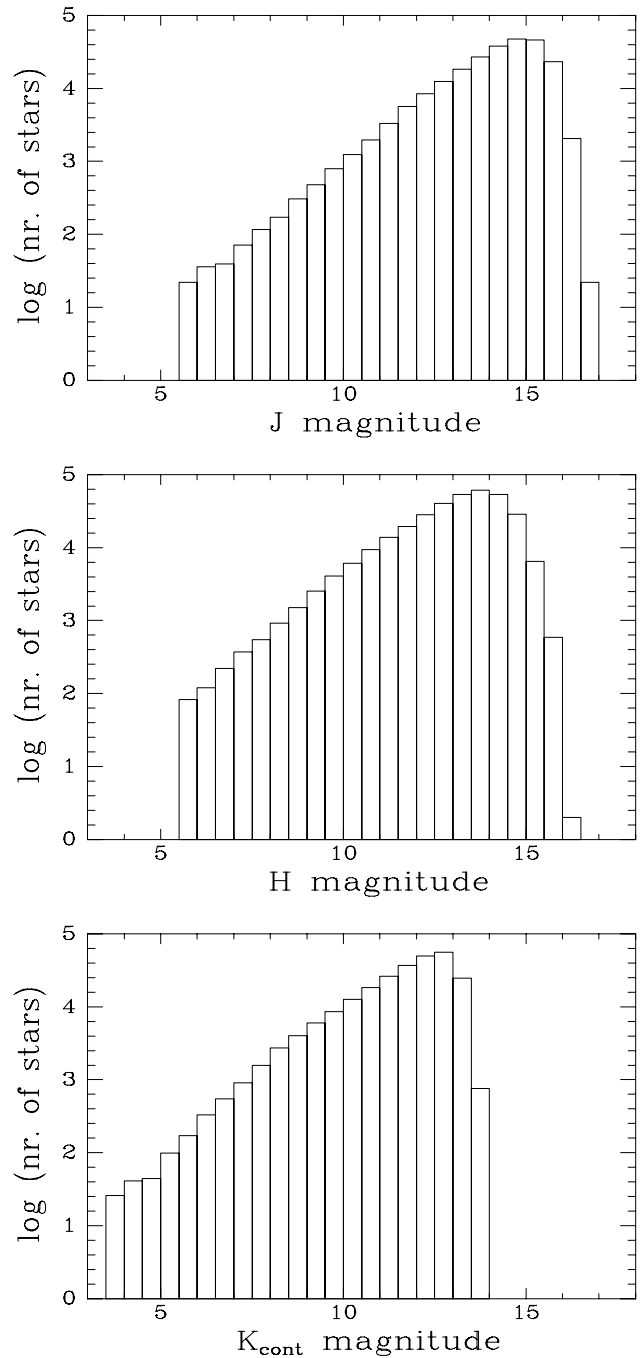}}
   \caption[]{Distribution of the magnitudes of the stars in our
sample in the $J$, $H$, and $K$-continuum filters. Only stars with
$\sigma < 0.3$~mag in the corresponding band are considered. The
sharp cutoff at bright magnitudes indicates the saturation limit.}
  \label{histo_mags}
\end{figure}

\begin{figure}
   \resizebox{8.5cm}{!}{\includegraphics{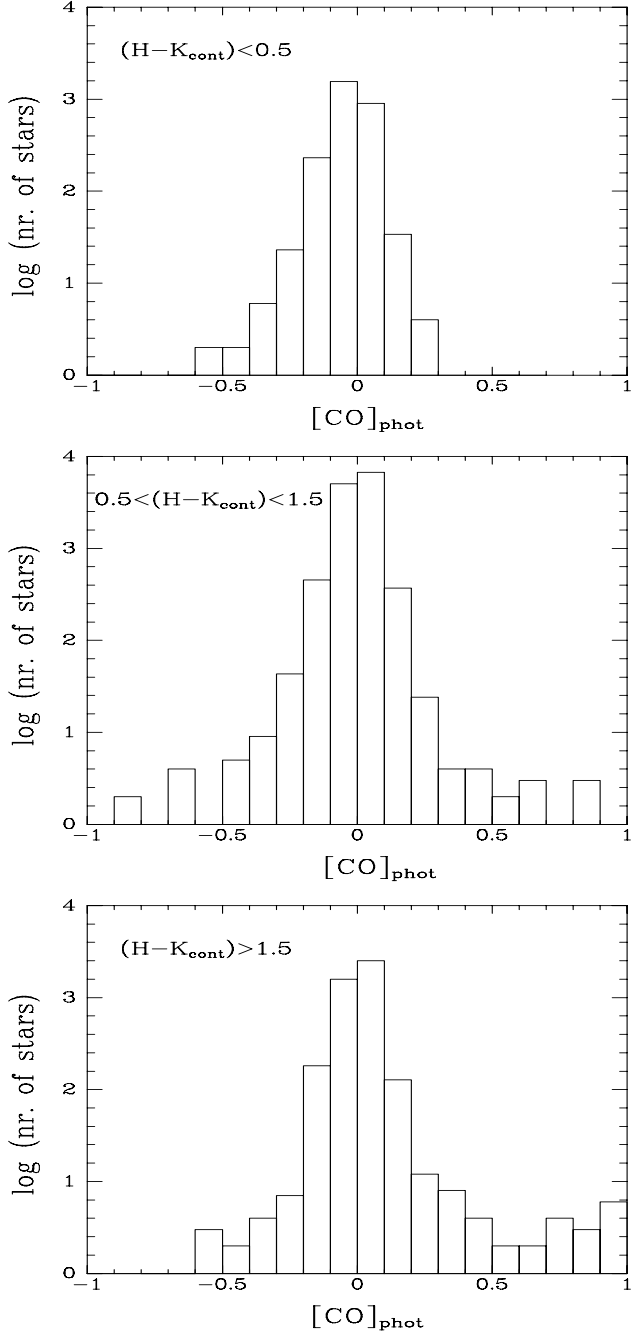}}
   \caption[]{The distribution of the ${\rm [CO]_{phot}}$ index
within three $(H-K)$ color bins, showing that cool stars are
abundant among the most reddened ones. Only stars with
$\sigma({\rm [CO]_{phot}}) < 0.1$ have been plotted here. Note
the use of a logarithmic scale in the vertical axis to enhance
the wings of the distribution of indices, which is strongly
peaked at ${\rm [CO]_{phot}} = 0$.}
  \label{histo_CO}
\end{figure}

\subsubsection {Properties of the photometrically selected sample}

  Our final catalog of detected sources contains 435,445 entries
corresponding to stars with their magnitudes determined to an
accuracy better than 0.3~mag (excluding the contribution of the
error in the zeropoint determination) in at least one of the $J$,
$H$ and $K$-continuum filters. This level is typically reached at
$J=15.5$, $H = 15.0$, $K = 13.0$. The shallower limit at $K$ is
due to the use of a narrow-band filter to estimate the $K$
magnitude. On the other hand, deeper limits in $J$ and $H$ are
needed to estimate the reddening towards the heavily obscured
stars that are the object of our program. Figure~\ref{histo_mags}
gives the histogram of source counts as a function of magnitude,
excluding stars with $\sigma > 0.3$~mag in each band. The ${\rm
[CO]_{phot}}$ can be usually measured down to an accuracy better
than 0.1~mag for stars down to $K \simeq 10.0$, as estimated from
the scatter of the measured ${\rm [CO]_{phot}}$ values as a
function of $K$. Although this is a rather modest limit in terms
of depth, as we will see it is sufficient for our primary purpose
of detecting late-type, distant supergiants. Our catalog contains
23,388 stars with their ${\rm [CO]_{phot}}$ indices determined to
that level of accuracy. These stars are plotted in
Figure~\ref{histo_CO} to illustrate the changing behavior of ${\rm
[CO]_{phot}}$ with the $(H-K)$ color index. The top panel shows a
sample with relatively blue $(H-K)$, which can be expected to be
dominated by nearby stars with low reddening. The distribution of
${\rm [CO]_{phot}}$ is approximately symmetric around a central
value of 0 with a very minor fraction of the stars having an index
significantly different from that value, in most cases due to
measurement errors caused by field crowding or the presence of
nearby companions. Such measurement errors can be reasonably
expected to affect stars regardless of their $(H-K)$. The second
panel, with moderately reddened stars, has a very similar
appearance but shows hints of skewness towards high values of
${\rm [CO]_{phot}}$. The skewness is more clear when we consider
the reddest stars in our sample, where we expect an increased
contribution of luminous late-type stars that can be detected even
at large distances and hence large amounts of foreground
reddening.

\subsubsection {Sample selection for spectroscopic follow-up}

  The imaging results described in the previous Section were then
used to select the best late-type cool star candidates to be
spectroscopically observed. This sample was composed of stars
having ${\rm [CO]_{phot}} > 0.15$ and a $K$-continuum magnitude
brighter than 10. Our selection was further refined by visually
inspecting the images in which the stars selected according to
these criteria appear, and discarding stars whose photometry was
uncertain due to proximity to the edge of the image (where
contamination by unfiltered bad pixels or cosmic rays was more
frequent) or confused by the proximity to nearby bright sources.
Such selection produced a list of 204 targets to be
spectroscopically observed, out of which three were discarded
because of the presence of a close companion that was not resolved
in the images from Calar Alto. The star numbers given in
Tables~\ref{total_sample}, \ref{tab_msg}, and
\ref{tab_msg_weakH2O} refer to the entry number in this list,
which is sorted by increasing right ascension. The positions and
broad-band photometry of the selected sources were obtained from
the 2MASS Point Source Catalog, as its astrometric and photometric
accuracy is better than that obtained for our sample with the
procedures described above.

\subsection{Near-infrared spectroscopy\label{obs_spectra}}

  The follow-up spectroscopy of the best candidate cool stars
selected as described above was obtained using the SOFI
near-infrared camera and spectrograph at the ESO New Technology
Telescope on La Silla, Chile, on the nights of 11, 12, and 13 July
2003. The setup and exposure parameters used were identical for
all the stars. We used the low-resolution red grism, which covers
most of the $H$ and $K$ bands from 1.53~$\mu$m to 2.52~$\mu$m at a
resolution $R = 980$ with the 0''6-wide slit that we used. Four
spectra were obtained for each star following an AB/BA pattern,
with each spectrum being the on-detector stack of three exposures
of 10 seconds each. The B5V star HD~169033, which lies near the
arc of the galactic plane where our targets are located, was
observed at periodic intervals of approximately one hour to sample
the evolution of the telluric absorption in the proximities of our
field. On the second night of our run we observed both HD~169033
and the solar analog HD~172411 (spectral type G3V) consecutively
and near their culmination, in order to minimize effects in the
telluric features affecting both.

  For the reduction of the spectra, we started by canceling out the
sky emission lines and the detector background by subtracting from
each spectrum taken on the A position another taken on the B
position, and viceversa. To correct for pixel-to-pixel variations
we used a flat field obtained with the telescope enclosure closed,
subtracting exposures taken with a continuum-emitting lamp
respectively on and off.

  The wavelength calibration was carried out in two steps. We first
obtained an exposure of a Xenon arc lamp, which produces a line
spectrum conveniently sampling the wavelength interval covered by
our observations, immediately after observing one of our targets
and keeping the instrument in the same position and setup, thus
eliminating any possible flexure effects or inaccuracies in
repositioning the slit. The arc lamp spectrum was
wavelength-calibrated and the dispersion solution was applied to
the sky spectrum of the object. In this way precise wavelenghts of
the sky lines and line blends at the resolution of our
observations were determined. The master calibrated sky spectrum
was then used to wavelength-calibrate the rest of the
observations. Each individual object spectrum was extracted, and
the same aperture trace was used to extract the corresponding sky
spectrum from an opposite frame in the AB/BA cycle. The lines
identified in the master sky spectrum were then reidentified in
the extracted sky spectrum at the position of the object and used
to determine a new dispersion solution, which was then applied to
the object spectrum. Finally, the four individual sky-subtracted,
flat-fielded, wavelength-calibrated spectra of each object were
combined into a single spectrum. An identical procedure was used
for the telluric calibration star and the solar analog
observations.

  To remove telluric features, the spectrum of each program star
was ratioed by the spectra of each of the telluric calibration
stars observed on the same night. The residual left after ratioing
at the position of the strong CO$_2$ features in the
2.00-2.05~$\mu$m interval was used as a quality criterion to
decide on the most suitable telluric calibration star to correct
for the telluric features in each program star. Since telluric
features generally depend on both time and airmass the selected
telluric calibration observation was often, but not always, the
one obtained closest in time to the observation of the program
star.

  The last step was relative flux calibration and removal
of photospheric features in the spectrum of the star used for
telluric correction. This was carried out by ratioing the
consecutive observations of the telluric calibrator and the solar
analog described earlier in this Section, and multiplying this
ratio by the flux-calibrated solar spectrum. The solar spectrum of
Livingston \& Wallace~(\cite{livingston91}) was used for this
purpose, with its resolution degraded to match that of our
observations, following the procedure described by Maiolino et
al.~(\cite{maiolino96}). The product of the telluric-over-solar
analog ratio and the solar spectrum thus produced a multiplicative
factor as a function of wavelength to be applied to our
telluric-calibrated program objects in order to obtain their
flux-calibrated spectrum in arbitrary units.

\begin{table*}
\caption{Position, 2MASS photometry, and equivalent widths of
selected features of all the observed stars}
\begin{tabular}{ccccccccccc} \hline
\noalign{\smallskip}
 & & & & & & \multicolumn{4}{c}{Equivalent widths} & \\
 Number  &      RA     &    dec      &  $K_S$ &
$J-H$ & $H-K_S$ &
CO(6,3) & NaI & CaI & CO(2,0) &$I({\rm H_2O})$ \\
 & (2000) & (2000) & \multicolumn{3}{c}{(2MASS)} & (\AA) & (\AA) &
 (\AA) & (\AA) & (mag) \\
\noalign{\smallskip} \hline \noalign{\smallskip}\\
001 &  17:58:58.83 & -23:43:03.7 & 10.36 &  2.11 & 0.87
& 5.65 & 4.34 &  3.71 & 15.64 &  -0.185 \\
002 & 17:59:17.61 & -23:48:02.3 &  9.93 &  2.09 & 0.92 &
8.34 & 4.04 & 3.25 & 21.00 &  -0.131 \\
003 & 17:59:19.99 & -23:49:34.8 &  9.86 &  1.23 & 0.53 &
2.34 & 2.31 & 2.70 & 11.33 &  -0.100 \\
004 & 17:59:20.46 & -23:51:11.7 &  9.53 & 2.47 & 1.14 &
9.45 & 5.11 & 3.42 & 21.99 &  -0.167 \\
005 & 17:59:20.63 & -23:34:01.1 & 9.64 & 2.21 & 0.95 &
7.84 & 4.32 & 3.59 & 20.44 & -0.180 \\
\noalign{\smallskip}\hline
\end{tabular}
\bigskip\\
Note: the full version of this Table is available from the CDS via
anonymous ftp to cdsarc.u-strasbg.fr (130.79.128.5)
\label{total_sample}
\end{table*}

  Of the 201 stars observed, 194 turned out to display obvious CO
absorption bands characteristic of spectral types K or M. We
discarded the 7 spectra of stars photometrically misclassified as
cool stars, as well as 3 additional spectra of faint stars showing
hints of CO absorption but having an insufficient signal-to-noise
ratio, which either prevented the determination of the equivalent
widths discussed in this paper or produced measurement errors not
comparable to those of the other stars in the sample.
Table~\ref{total_sample} (available in full electronically only
\footnote{These data in electronic form can be accessed from the
CDS via anonymous ftp to cdsarc.u-strasbg.fr (130.79.128.5)})
contains the entire sample of 191 stars presented in this paper
listing their positions, 2MASS photometry, and measured equivalent
widths of selected features (see Section~\ref{eqw}). The numbering
of the objects discussed in this paper refers to the entry number
in that Table. The complete collection of reduced spectra used in
the forthcoming discussions is also available electronically.

  As an additional step, we computed radial velocities for all
our program stars by cross-correlating their $K$-band spectrum to
a zero-velocity cool star template spectrum obtained by combining
the spectra of M-type giant and supergiant stars in the atlas of
Wallace \& Hinkle (\cite{wallace97}). Doppler-corrected spectra
were produced in this way for all our program stars, as well as
for the stars in the Lan\c{c}on \& Woods (\cite{lancon00}) atlas
that we use extensively as a reference in the present work (see
Section~\ref{ref_sample}). The Doppler-corrected spectra were the
ones used for the derivation of the equivalent widths discussed in
this paper. The accuracy of the radial velocities obtained is
limited given the resolution of our spectra, and we estimate it to
be approximately 30-40~km~s$^{-1}$ from the residuals of the
dispersion solutions and the width of the cross-correlation peaks.

\begin{figure*}
   \resizebox{13cm}{!}{\includegraphics{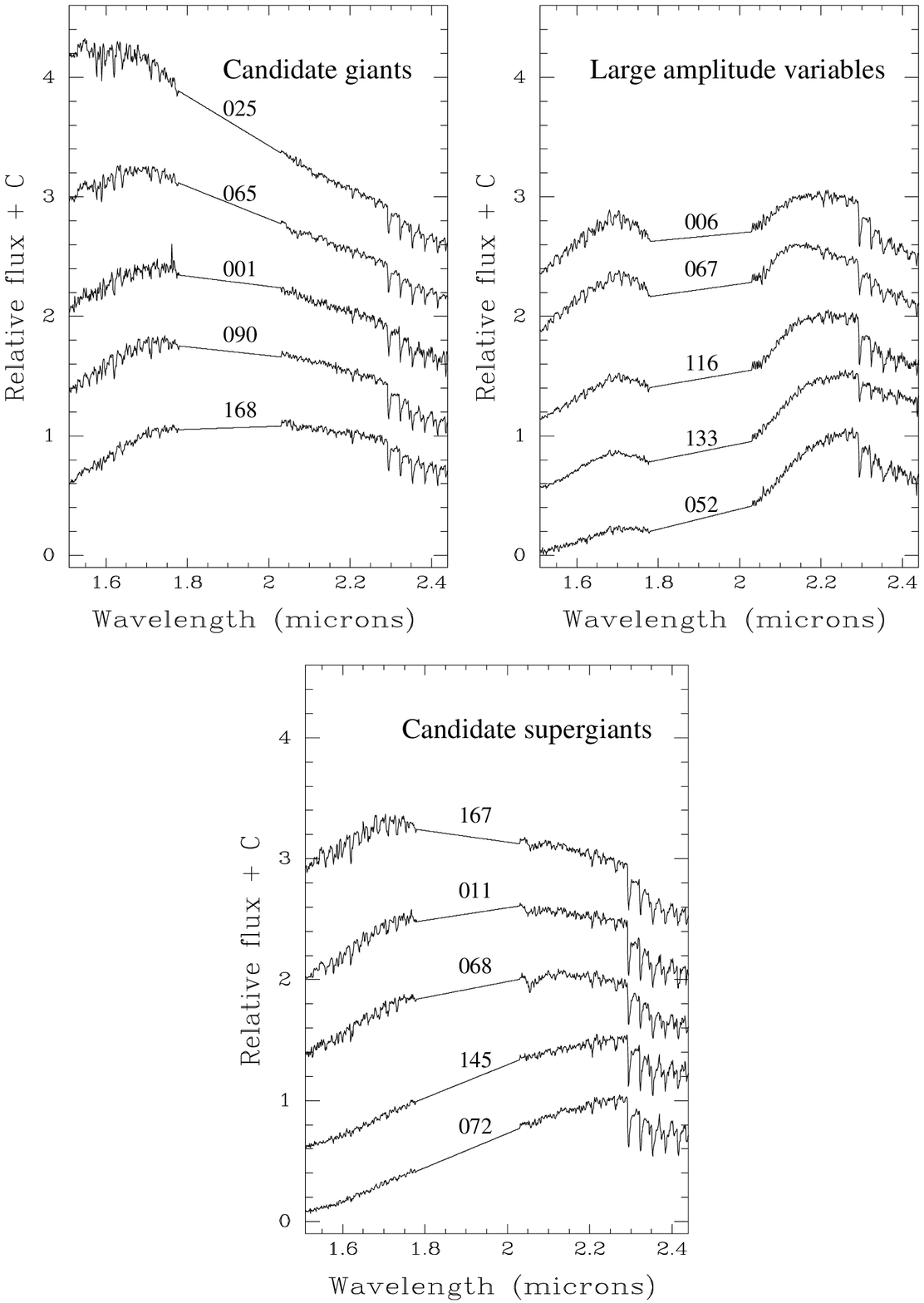}}
   \caption[]{Representative examples of cool stellar spectra
found in our sample. The interval between 1.80~$\mu$m and
2.05~$\mu$m is contaminated by strong telluric absorption bands
and has been clipped off. The spectra in the top left panel show
relatively strong CO absorption bands, but no signs of depression
due to the wings of the broad water steam feature centered near
1.9~$\mu$m. The top right panel shows a sample of stars with
similarly strong CO, but now also with a conspicuous H$_2$O
feature that we classify as large amplitude variables following
LW2000. The stars in the bottom set of spectra are drawn from the
sample of candidate supergiant stars (see
Section~\ref{cand_supergiants}). The spectra are normalized to the
flux between 2.20 and 2.25~$\mu$m, and the offsets between
consecutive spectra are of 0.5 in these normalized flux units.The
number accompanying each spectrum refers to the entry number in
Table~\ref{total_sample}.}
  \label{samplespec}
\end{figure*}

\section{Results\label{results}}

\subsection {Equivalent widths and spectrophotometric indices\label{eqw}}

  At the relatively low resolution available, the main features
that are visible in our $K$-band spectra are the CO bandheads
redwards of 2.293~$\mu$m, as well as the NaI feature at
2.208~$\mu$m and the CaI feature at 2.264~$\mu$m\footnote{These
features receive their names from the atomic species that normally
dominate the absorption at those wavelengths in spectra with
resolutions $R \sim 1000$ or lower. However, higher resolution
observations reveal contributions also by CN and atomic species
that can become important below 4,000~K (Ram\'\i rez et
al.~\cite{ramirez97}), as discussed in Section~\ref{discussion}.
Although in this paper we refer to these features by their usual
denomination in the literature (NaI, CaI), the reader should be
aware that these species may not be the only ones contributing to
their strength.}. Other atomic lines such as those of FeI and MgI
are also easily visible, as well as weaker features due to CN.
More features appear in the $H$ band. The reader is referred to
Kleinmann \& Hall (\cite{kleinmann86}), Wallace \& Hinkle
(\cite{wallace97}) and Hinkle et al. (\cite{hinkle95}) for
detailed line identifications in these two bands. The overall
shape of the continuum is strongly affected by the presence or
absence of the broad wings of the water absorption band centered
around 1.9~$\mu$m, where the separation between the $H$ and $K$
bands lies. Figure~\ref{samplespec} shows some selected spectra
representative of the variety found in our sample.

  A considerable number of slightly different definitions of spectral
passbands useful for the quantification of different features can
be found in the literature (e.g. Lan\c{c}on \&
Rocca-Volmerange~\cite{lancon92}; Ram\'\i rez et
al.~\cite{ramirez97}; Lan\c{c}on \& Wood \cite{lancon00};
F\"orster-Schreiber \cite{forster00}; Frogel et al.
\cite{frogel01}), which are normally defined in the most
convenient way given the spectral resolution available. Our choice
has been to adopt the $K$-band indices defined by Frogel et
al.~(\cite{frogel01}) for the CaI, NaI, and CO(2,0) features,
which are also directly comparable to those used by Ram\'\i rez et
al.~(\cite{ramirez97}, \cite{ramirez00b}). They thus allow a
straightforward comparison between the spectral characteristics of
our objects and those of the extensive samples of RGB stars
observed by those authors in the solar neighbourhood (Ram\'\i rez
et al.~\cite{ramirez97}), the galactic bulge (Ram\'\i rez et
al.~\cite{ramirez00b}), and globular clusters (Frogel et
al.~\cite{frogel01}) (see Section~\ref{ref_sample}). Moreover, we
have defined additional passbands to measure SiI and CO(6,3)
features in the $H$ band, and FeI, MgI in the $K$ band. In all
cases except for the $K$-band CO features the equivalent widths
are measured with respect to the local continuum level
interpolated between two featureless points on both sides of the
feature. The lack of a continuum longwards of 2.29~$\mu$m in our
spectra forces us to use an extrapolation instead using two
continuum points at shorter wavelengths. The definition of the on-
and off-feature bands used in this paper is given in
Table~\ref{passbands}. Note that the interpolation or
extrapolation using two continuum bands makes the measured
equivalent widths virtually insensitive to reddening, which is not
the case for the [CO]$_{\rm phot}$ defined earlier. Typical
uncertainties in the equivalent widths that we measure in our
spectra are $\pm 0.5$~\AA, except for the CO(2,0) band where we
estimate $\pm 1.0$~\AA\ due to the need to extrapolate from the
bluer continuum.

\begin{table}
\caption{Feature passbands and their reference continua}
\begin{tabular}{ccccccc} \hline
\noalign{\smallskip}
 & \multicolumn{2}{c}{line/band} & \multicolumn{2}{c}{blue cont.} & \multicolumn{2}{c}{red cont.} \\
Feature & $\lambda_0$ & $\Delta \lambda$ & $\lambda_0$ & $\Delta
\lambda$ & $\lambda_0$ & $\Delta \lambda$ \\
 & ($\mu$m) & ($\mu$m) & ($\mu$m) & ($\mu$m) & ($\mu$m) &
 ($\mu$m)\\
\noalign{\smallskip} \hline \noalign{\smallskip}
SiI & 1.5898 & 0.0035 & 1.587 & 0.002 & 1.5953 & 0.0025 \\
CO(6,3) & 1.620 & 0.0117 & 1.6140 & 0.006 & 1.6325 & 0.009 \\
CN  & 2.1335 & 0.005 & 2.127 & 0.008 & 2.139 & 0.006 \\
NaI & 2.2075 & 0.007 & 2.194 & 0.006 & 2.215 & 0.004 \\
FeI & 2.2275 & 0.006 & 2.2145 & 0.006 & 2.2327 & 0.0045 \\
CaI & 2.2635 & 0.011 & 2.2505 & 0.011 & 2.271 & 0.002 \\
MgI & 2.2807 & 0.0035 & 2.277 & 0.004 & 2.2865 & 0.008 \\
CO(2,0) & 2.2955 & 0.013 & 2.250 & 0.016 & 2.2875 & 0.007 \\
\hline\smallskip
\end{tabular}
{\it Definition of passbands for measurement of H$_2$O band
wings\smallskip\\}
\begin{tabular}{ccc}\hline
\noalign{\smallskip}
& $\lambda_0$ & $\Delta \lambda$ \\
 & ($\mu$m) & ($\mu$m) \\
\noalign{\smallskip} \hline \noalign{\smallskip}
$f^C_H$ & 1.675 & 0.05 \\
$f^L_H$ & 1.755 & 0.05 \\
$f^L_K$ & 2.075 & 0.05 \\
$f^C_K$ & 2.255 & 0.05 \\
\hline
\end{tabular}
\bigskip\\
Note: passbands for the measurement of the NaI, CaI, and CO(2,0)
features are defined as in Frogel et al.~(\cite{frogel01}) and
Schultheis et al.~(\cite{schultheis03}). \\
\label{passbands}
\end{table}

\begin{figure}
   \resizebox{8.5cm}{!}{\includegraphics{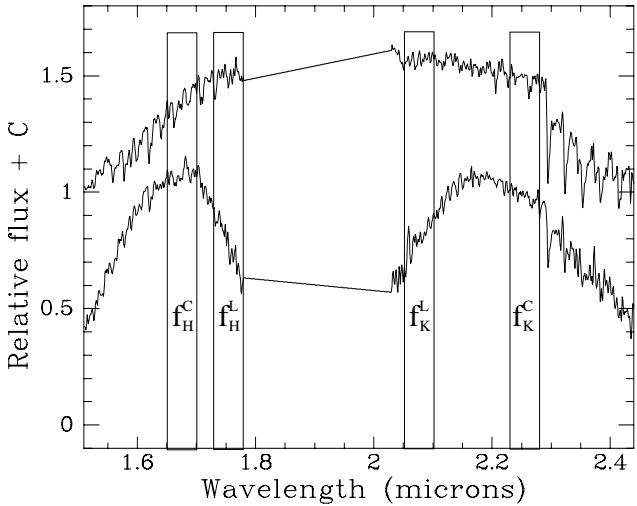}}
   \caption[]{The positions of the spectrophotometric bands defined in
Section 3.1, plotted on representative spectra from our sample to
show their usefulness in measuring the flux depression caused by
strong water band wings.}
  \label{h2oindex}
\end{figure}

  The width of the wings of the H$_2$O feature separating the $H$
and $K$ bands requires a definition different from the equivalent
widths discussed above to quantify their strength. We have thus
defined a reddening-free water index, $I({\rm H_2O})$, based on
the comparison of the fluxes in two bands located on the blue and
red wing of the feature and the approximate continuum level
measured on unaffected parts of the $H$ and $K$ bands. The
locations and widths of these bands are also listed in
Table~\ref{passbands}, where $f^L_H$, $f^L_K$ denote the fluxes on
the $H$-band (blue) and $K$-band (red) wing of the feature, and
$f^C_H$, $f^C_K$ are the respective continuum fluxes.
Figure~\ref{h2oindex} shows the position of the bands used to
measure $I({\rm H_2O})$. Assuming that the extinction has a
wavelenght dependence of the form $A_\lambda \propto \lambda ^
{-1.7}$ (Mathis \cite{mathis90}), the quantity

$$I({\rm H_2O}) = -2.5 \log {f^L_H \over f^C_H}
  - 2.081 \log {f^L_K \over f^C_K}  \eqno (1)$$

\noindent is thus a reddening-free measurement of the depression
in flux due to water vapour absorption on both sides of the gap
separating the $H$ and $K$ bands, which we express in magnitudes
for convenience.

\subsection{The reference samples\label{ref_sample}}

\begin{figure}
   \resizebox{8.5cm}{!}{\includegraphics{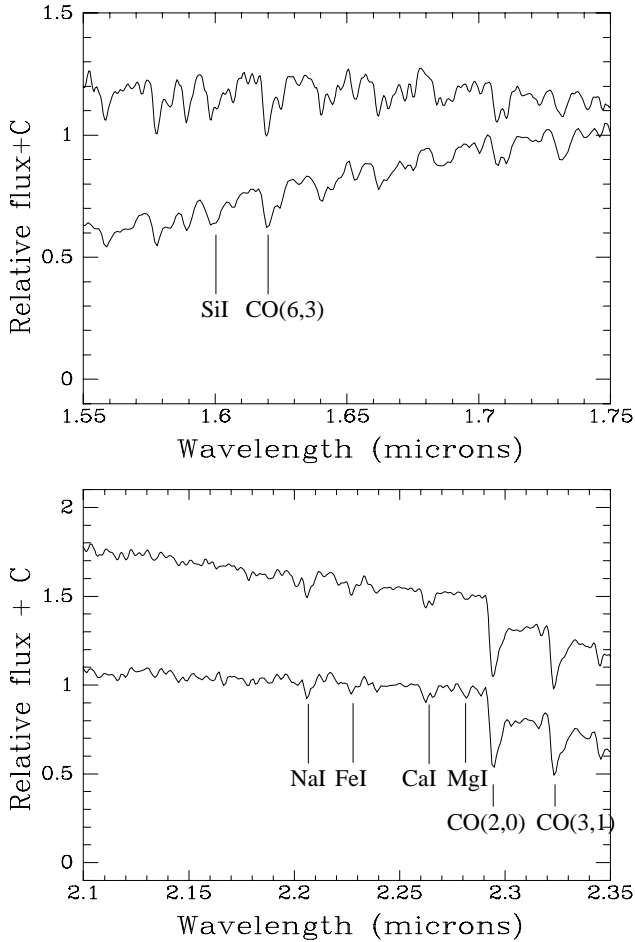}}
   \caption[]{A comparison between the $H$ and $K$ spectra of
one supergiant from the LW2000 library of spectra (above) and one
supergiant candidate from our sample (below), showing their
similar resolution and quality. The comparison also shows the
reality of most of the features that can be seen in our spectra.
The difference in continuum slopes, especially in the $H$ band, is
due to the higher extinction towards the star of our sample. The
features listed in Table~\ref{passbands} are noted.}
  \label{comp_lancon}
\end{figure}

  Low- and medium-resolution atlases of cool stars in the infrared
have been presented in a number of studies (e.g. Kleinmann \&
Hall~\cite{kleinmann86}, Lan\c{c}on \& Rocca-Volmerange
\cite{lancon92}, Dallier et al.~\cite{dallier96}, Wallace \&
Hinkle~\cite{wallace96}, \cite{wallace97}, Meyer et
al.~\cite{meyer98}, F\"orster-Schreiber~\cite{forster00}, Ivanov
et al.~\cite{ivanov04}; see this latter reference for a more
exhaustive list of relevant work), with the general twofold
purpose of providing a tool both for the classification of
individual objects and for the construction of templates of
integrated spectral energy distributions for population synthesis
analyses. Among these atlases, a particularly suitable one for the
investigation of the properties of our sample is provided by the
extensive collection of cool stellar spectra in the visible and
the near infrared published by Lancon \& Wood (\cite{lancon00},
hereafter LW2000). Their sample provides a wide variety of spectra
corresponding to different classes of luminous cool stars in
different environments such as the solar neighbourhood, the
galactic bulge, and the Magellanic Clouds, and includes multiepoch
observations for a large number of variable stars. Its resolution,
$R = 1100$, closely matches that of the spectra presented in this
paper, thus facilitating a direct comparison (see
Figure~\ref{comp_lancon}). Classes represented in LW2000 include
supergiants, thermally pulsating AGB stars of high and low
amplitudes, non-pulsating RGB stars, and carbon stars.

  We have used the spectra from LW2000 for which those authors
present observations in the $H$ and $K$-band and belonging to the
samples of supergiants, oxygen-rich variables, bulge giants, and
non-pulsating RGB stars. Unfortunately, few spectra in the $H$ and
$K$ bands are available for the latter class of objects in the
LW2000 sample. This shortcoming can be largely circumvented by
adding the collection of equivalent widths of the NaI, CaI, and
CO(2,0) features in the $K$-band spectrum of bright giants
measured by Ram\'\i rez et al.~(\cite{ramirez97}) on spectra
having a resolution similar to ours. We have also used the lists
of equivalent widths of the same features in Ram\'\i rez et
al.~(\cite{ramirez00b}) for as sample of over 100 bulge RGB stars,
and of RGB stars in the most metal-rich globular clusters observed
by Frogel et al.~(\cite{frogel01}), namely \object{NGC~5927}
(${\rm [Fe/H]} = -0.37$), \object{NGC~6440} (${\rm [Fe/H]} =
-0.34$), \object{NGC~6528} (${\rm [Fe/H]} = -0.17$), and
\object{NGC~6553} (${\rm [Fe/H]} = -0.25$). Low-metallicity
objects such as those in the Magellanic Clouds sample of LW2000
are not expected to be significantly represented among our inner
galactic disk objects, and have not been considered here.
Likewise, there are clearly no carbon-rich stars in our sample
(which would be easily distinguishable thanks to the strong C$_2$
feature at 1.77~$\mu$m) and we have thus excluded the objects of
that class in LW2000 atlas from the comparison with our results.

\subsection{The H$_2$O vs. CO diagrams\label{H2O_CO}}

  A first classification of a sample of cool luminous star can be
established on the basis of the relative importance of the most
prominent molecular features in the $H$ and $K$ bands due to water
steam and carbon monoxide. CO is to a first approximation a
temperature indicator with the depth of the bands increasing as
temperature decreases. However, the depth of the CO bandhead at
2.293~$\mu$m is also sensitive to the surface gravity and is often
used as a discriminator between giants and supergiants at a given
temperature (e.g. Kleinmann \& Hall~\cite{kleinmann86},
F\"orster-Schreiber~\cite{forster00}, Ivanov et
al.~\cite{ivanov04}). The CO band intensity is also sensitive to
the carbon abundance in the atmospheres of oxygen-rich stars and
is thus an indicator of the importance of the dredge-up processes
in the post-main sequence evolution of the star (Carr et
al.~\cite{carr00}). On the other hand, strong H$_2$O absorption is
formed in the cool, dense outer layers of large amplitude
Mira-type and semiregular variables possessing very extended
atmospheres with a complicated structure determined by shocks
propagating through them (Bessell et al.~\cite{bessell89} and
references therein), and can be used as a discriminator between
pulsating AGB stars and supergiants (Lan\c{c}on \&
Rocca-Volmerange~\cite{lancon92}), both of which have overlapping
luminosity ranges.

\begin{figure}
   \resizebox{8.5cm}{!}{\includegraphics{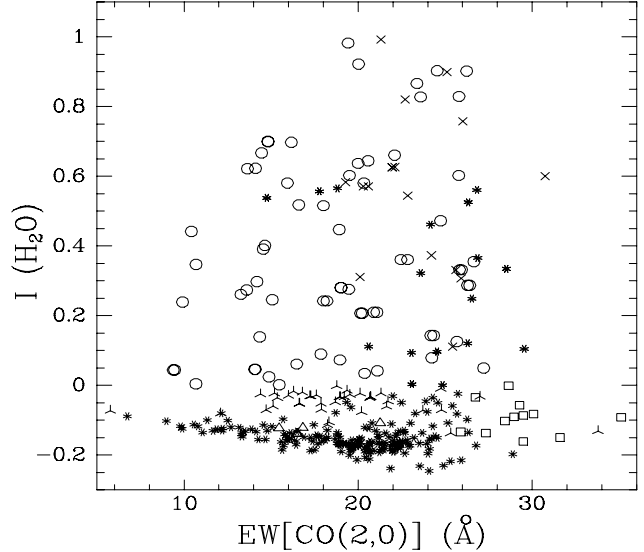}}
   \caption[]{$EW[{\rm CO}(2,0)]$ vs. $I({\rm H_2O})$ diagram for all
the stars in our sample (eight-pointed asterisks), compared to the
different classes of cool stars of LW2000: non-pulsating red
giants (triangles), supergiants (squares), bulge giants (crosses),
and oxygen-rich variables, separated between those having strong
(circles) and weak (three-pointed asterisks) water absorption.}
  \label{CO20_H2O}
\end{figure}

  Figure~\ref{CO20_H2O} shows the distributions of our
sample and the LW2000 one in the CO(2,0)-H$_2$O diagram. The
apparent overabundance of objects with strong water absorption in
LW2000 is largely due to the fact that we have plotted in the
diagram every single observation of each object in LW2000 as a
separate point; therefore, any given star observed at different
phases by LW2000 appears multiple times in the diagram.
Figure~\ref{CO20_H2O} shows that our sample is strongly dominated
by objects with the spectral characteristics of non-pulsating
stars that occupy the {\it locus} of red giants and supergiants.
The small scatter of $I({\rm H_2O})$ for objects spanning a wide
range of extinctions confirms that the extinction law chosen to
derive Eq.~(1) yields indeed a reddening-free index, whereas its
steady decline between $EW[{\rm CO(2,0)}] = 10$~\AA\ and 20~\AA\
is due to the progressive departure of the $H$ and $K$ band
spectrum from a Rayleigh-Jeans shape as the temperature decreases
and the depth of the CO bands increases. Most of our objects have
moderate values of $EW[{\rm CO(2,0)}]$ that are generally well
below those of the sample of LW2000 supergiants, a feature that we
will consider in the coming Sections. Finally, the objects with
strong water absorption have a preference towards high $EW[{\rm
CO(2,0)}]$, unlike in the LW2000 sample where large positive
$I({\rm H_2O})$ indices are present over the whole range of
$EW[{\rm CO(2,0)}]$ considered. The trend in our sample is
qualitatively similar to that observed in Figure~13 of Lan\c{c}on
\& Rocca-Volmerange~(\cite{lancon92}) where their objects split
along two branches (corresponding to RGB stars and supergiants) at
the highest $EW[{\rm CO(2,0)}]$, although we do not find the gap
between both branches suggested by the results of those authors.
Unfortunately, no information on the H$_2$O depression is
available in the data published by Ram\'\i rez et
al.~(\cite{ramirez97}) given their wavelength coverage restricted
to the red part of the $K$ band.

\begin{figure}
   \resizebox{8.5cm}{!}{\includegraphics{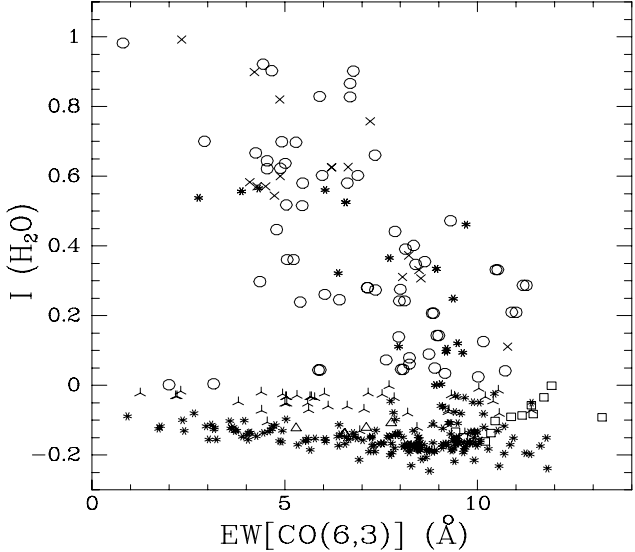}}
   \caption[]{$EW[{\rm CO(6,3)}]$ vs. $I({\rm H_2O})$ diagram
comparing the stars in our sample with those of LW2000. Symbols
are as in Figure~\ref{CO20_H2O}.}
  \label{CO63_H2O}
\end{figure}

  Figure~\ref{CO63_H2O} is the analogous of Figure~\ref{CO20_H2O}, now
replacing the CO(2,0) band by the CO(6,3) band at 1.62~$\mu$m.
Again cool supergiants of the LW2000 sample occupy the region of
highest CO absorption, now with a larger overlap with our sample.
A difference with Figure~\ref{CO20_H2O} is that objects with
$I({\rm H_2O}) > 0$ now populate a broad but clearly defined band
so that the objects with the stronger water absorption tend to
have the lowest $EW[{\rm CO(6,3)}]$. The same behaviour in this
respect can be seen in the LW2000 sample and ours. A closer
examination shows that not only the CO(6,3) band, but all the
features in the $H$ band are weaker among the stars with strong
water band wings, indicating the existence of significant veiling
correlated with the strength of the water feature. Some degree of
veiling in the $K$ band as well may be responsible for the
relatively weak CaI and NaI features of stars with $I({\rm H_2O})
> 0$ in Figures~\ref{CaI_CO} and \ref{NaI_CO}.

\subsection{The NaI and CaI features \label{atomic}}

  Diagrams combining absorption in atomic lines and molecular
bands having different dependencies on temperature, surface
gravity, and metallicity offer in principle important diagnostics
on the intrinsic properties of samples of individual cool stars
and of composite populations, and have been discussed in a number
of studies (see Ivanov et al.~\cite{ivanov04} for a recent,
comprehensive review). A comparison between our sample in the
inner Galaxy, the sample of LW2000, and the samples of RGB stars
in the solar neighbourhood, the bulge, and globular clusters
described in Section~\ref{ref_sample} can provide insight on
possible intrinsic differences of both populations and the way
they reflect themselves on measurable spectral features.

\begin{figure}
  \resizebox{8.5cm}{!}{\includegraphics{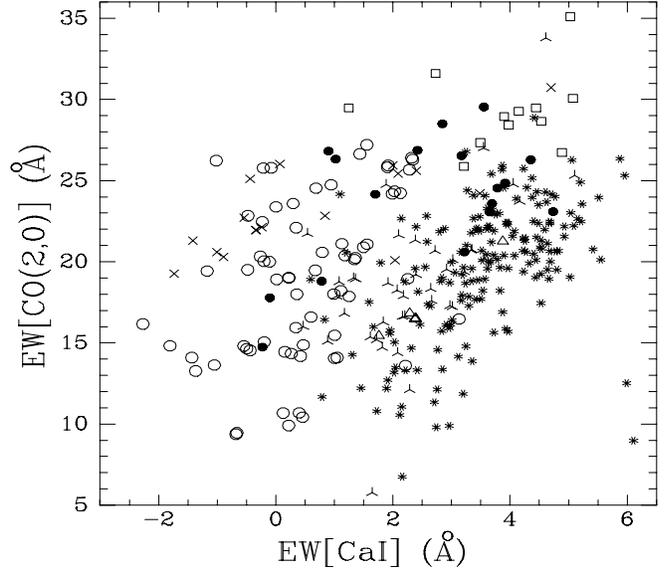}}
  \caption[]{$EW[{\rm CaI}]$ vs. $EW[{\rm CO(2,0)}]$ diagram
comparing the stars in our sample with those of LW2000. Symbols
are in general as in Figure~\ref{CO20_H2O}, but now the stars of
our sample with $I({\rm H_2O}) > 0$ are marked as filled circles.
Note the offset between the mean distribution of stars of our
sample and other reference samples.}
  \label{CaI_CO}
\end{figure}

\begin{figure}
   \resizebox{8.5cm}{!}{\includegraphics{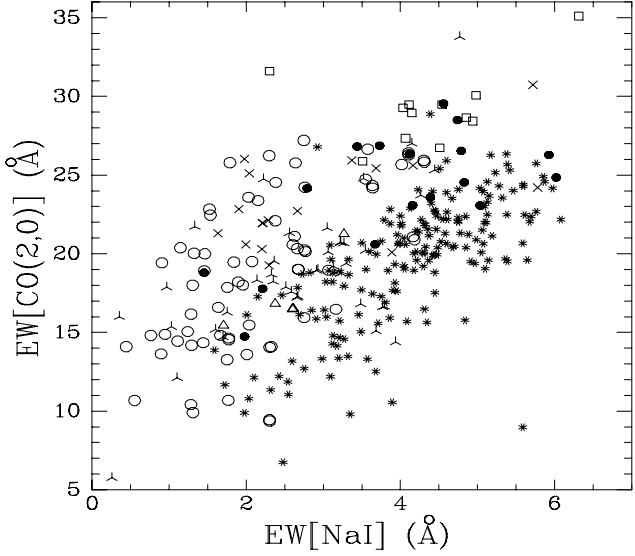}}
   \caption[]{$EW[{\rm NaI}]$ vs. $EW[{\rm CO(2,0)}]$ diagram
comparing the stars in our sample with those of LW2000. Symbols
are as in Figure~\ref{CaI_CO}. Like in Figure~\ref{CaI_CO}, there
is a clear offset between the mean distribution of stars of our
sample and that of other reference samples. Note that a large
fraction of our stars reach $EW[{\rm NaI}]$ values higher than
those of most local supergiants.}
  \label{NaI_CO}
\end{figure}

\begin{figure}
   \resizebox{8.5cm}{!}{\includegraphics{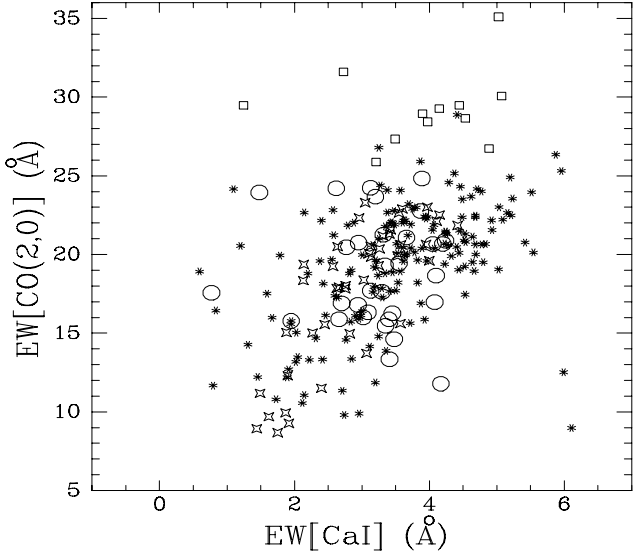}}
   \caption[]{
Same as Figure~\ref{CaI_CO}, but now comparing our sample to that
of RGB stars in the solar neighbourhood from Ram\'\i rez et
al.~(\cite{ramirez97}; four-pointed stars) and in relatively
metal-rich globular clusters from Frogel et al.~(\cite{frogel01})
(open circles). The LW2000 sample of supergiants (open squares as
in previous figures) is also included.}
  \label{CaI_CO_ramirez97}
\end{figure}

\begin{figure}
   \resizebox{8.5cm}{!}{\includegraphics{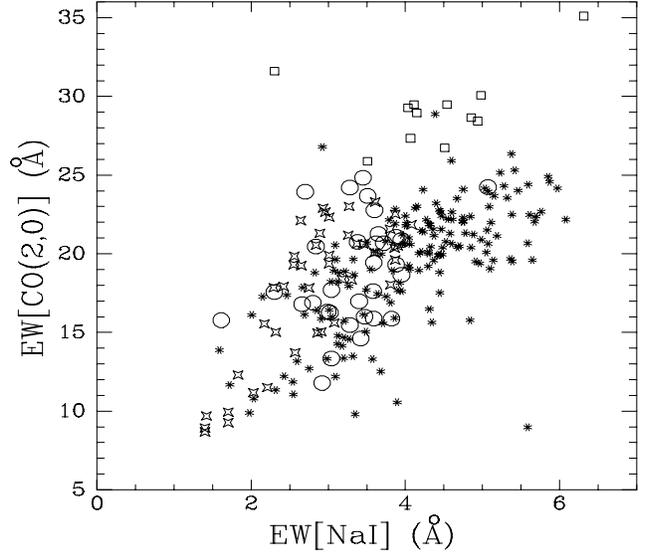}}
   \caption[]{
Same as Figure~\ref{NaI_CO}, but now comparing our sample to that
of RGB stars in the solar neighbourhood from Ram\'\i rez et
al.~(\cite{ramirez97}; four-pointed stars) and in relatively
metal-rich globular clusters from Frogel et al.~(\cite{frogel01})
(open circles). The LW2000 sample of supergiants (open squares as
in previous figures) is also included. Note that most of our stars
have $EW[{\rm NaI}]$ greater than those of the coolest RGB
globular cluster stars.}
  \label{NaI_CO_ramirez97}
\end{figure}

\begin{figure}
   \resizebox{8.5cm}{!}{\includegraphics{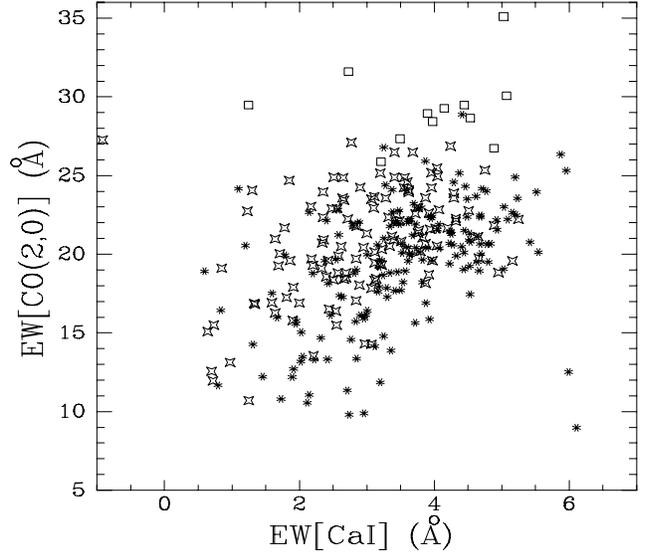}}
   \caption[]{
Same as Figure~\ref{CaI_CO}, but now comparing our sample to that
of RGB stars in the galactic bulge from Ram\'\i rez et
al.~(\cite{ramirez00b}; four-pointed stars). The LW2000 sample of
supergiants (open squares as in previous figures) is also
included.}
  \label{CaI_CO_ramirez00}
\end{figure}

\begin{figure}
   \resizebox{8.5cm}{!}{\includegraphics{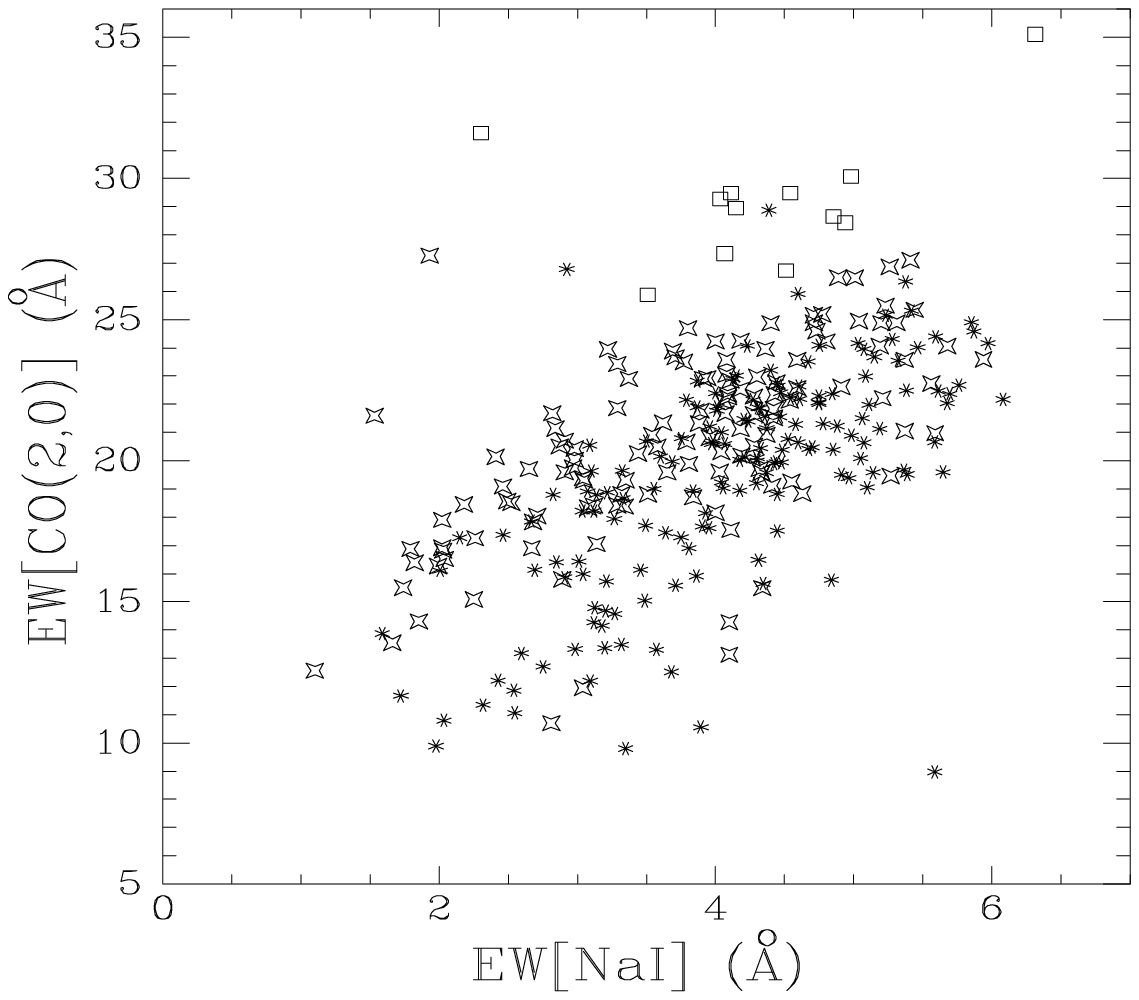}}
   \caption[]{
{Same as Figure~\ref{NaI_CO}, but now comparing our sample to that
of RGB stars in the galactic bulge from Ram\'\i rez et
al.~(\cite{ramirez00b}; four-pointed stars). The LW2000 sample of
supergiants (open squares as in previous figures) is also
included}. Note that most of our stars have $EW[{\rm NaI}]$
greater than those of the coolest RGB globular cluster stars.}
  \label{NaI_CO_ramirez00}
\end{figure}

  We have plotted in Figures~\ref{CaI_CO} and \ref{NaI_CO} the
relationship between the NaI and CaI features and the CO(2,0)
bandhead. In particular, the combination of the strength of a
feature essentially dependent on the temperature such as CaI and
one that combines dependence on both temperature and surface
gravity such as CO(2,0) has been proposed as a luminosity class
discriminator (F\"orster-Schreiber~\cite{forster00}, Ivanov et
al.~\cite{ivanov04}).

  Figures~\ref{CaI_CO} and \ref{NaI_CO} do not show tight
overall correlations for the LW2000 sample, mainly due to the
abundance in it of AGB stars whose complex, time-dependent
atmospheric structure is expected to blur them. However, local
supergiants dominate in a well defined region of the diagram
characterized by the strongest absorptions in both the CaI and NaI
features and CO(2,0).

  The loose correlation between the CaI or NaI features and
CO(2,0) is much more apparent when only RGB stars and supergiants
are considered, as shown in Figures~\ref{CaI_CO_ramirez97} to
\ref{NaI_CO_ramirez00}, where our sample is compared to the local
supergiants of LW2000 and the RGB stars in the solar neighbourhood
(Ram\'\i rez et al.~\cite{ramirez97}), the galactic bulge (Ram\'\i
rez et al.~\cite{ramirez00b}), and the relatively metal-rich
globular clusters of Frogel et al.~(\cite{frogel01}). Clear
systematic difference are noted between the stars in our sample on
one side, and the local RGB sample of Ram\'\i rez et
al.~(\cite{ramirez97}) and the sample of RGB stars in globular
clusters of Frogel et al.~(\cite{frogel01}) on the other, in the
sense that the equivalent widths of the atomic features in our
sample extend to higher values. Local RGB stars occupy essentially
the same region of the diagrams presented in
Figures~\ref{CaI_CO_ramirez97} and \ref{NaI_CO_ramirez97} as RGB
stars in globular clusters, which are also plotted in those
figures. This is especially clear in the NaI vs. CO(2,0) diagram,
where the overlap in $EW{\rm [NaI]}$ between the giants and
supergiants of the reference samples is minor, and where only one
RGB star belonging to \object{NGC~6528} (the highest metallicity
cluster in Frogel et al.~(\cite{frogel01}) sample) appears with
$EW{\rm [NaI]} > 4.0$~\AA. The strengths of the CO(2,0) bandhead
in our stars are otherwise similar to, or only slightly exceeding,
the ones of the coolest RGB stars in the local and globular
cluster samples. On the other hand, the galactic bulge sample has
overall characteristics much closer to those of our objects, as
shown in Figures~\ref{CaI_CO_ramirez00} and
\ref{NaI_CO_ramirez00}. The largest measured equivalent widths of
the CO(2,0) feature are similar to those of the local RGB star
sample, which may be explained by their proximity to saturation
and subsequent insensitivity to metallicity differences. However,
RGB stars in the bulge do reach the high CaI and NaI equivalent
widths that we measure in our sample and their {\it loci} in those
diagrams has a much broader, although not complete, overlap with
the one of our sample. The overlap with this sample is
nevertheless poorer at low equivalent widths, where the local RGB
stars from Ram\'\i rez et al.~(\cite{ramirez97}) rather than bulge
ones tend to populate the region occupied by the stars of our
sample. This is to be expected since stars with lower equivalent
widths are hotter, and therefore intrinsically fainter and more
nearby on the average as indicated by the lower reddening of their
colors, and can thus be considered as an extension of the local
sample.

  Metallicity differences between RGB stars in the solar
neighbourhood and in the bulge are the most obvious candidate to
produce the systematic shifts between samples in the equivalent
width diagrams. The quantification of these differences from low
resolution spectra is however difficult, as the use of
relationships calibrated on globular clusters implies their
extrapolation outside their strict range of applicability
(temperatures, luminosities, metallicities, feature strengths,
star formation and chemical pollution environments...) and even
become dependent on the functional form adopted for the
relationships, as illustrated by Figure~6 of Ram\'\i rez et
al.~(\cite{ramirez00b}). Moreover, the low resolution spectrum may
be also determined by other effects that mimic metallicity
effects, as discussed in Section~\ref{gal_cen}.

\subsection{Color-magnitude diagram and luminosities\label{col_mag}}

  The obvious differences in the intrinsic properties of the
local sample of giants and the ones in the inner galactic disk,
represented by both our objects and the bulge sample of Ram\'\i
rez et al.~(\cite{ramirez00b}), leads us to be cautious about
possible differences also existing between local supergiants and
those in the inner Galaxy that may render inapplicable the
spectroscopic criteria allowing a separation between giants and
supergiants in the solar neighbourhood. Indeed, such criteria do
not seem to apply at least to giants and supergiants at the
galactic center, as concluded by Schultheis et
al.~(\cite{schultheis03}) from their near-infrared study of a
sample of cool, luminous stars detected in the ISOGAL survey
(Omont~\cite{omont96}) with spectra obtained with the same
instrumentation as ours. Schultheis et al.'s identification of
four very likely cool supergiants rests on their derived
luminosities, which they estimate based on the assumption that all
the sources in their sample lie at the distance of the galactic
center. This is a reasonable working hypothesis in their case,
given the high density of sources detected within a radius of less
than $1^\circ 5$ from the galactic center in which foreground
contamination is expected to be negligible. Unfortunately we
cannot make a similar assumption on the distance, and are thus
forced to examine more indirect means of estimating the luminosity
of our stars in order to assess the possible presence of
supergiants among them, without relying on spectroscopic criteria
derived from the solar neighbourhood.

\begin{figure}
   \resizebox{8.5cm}{!}{\includegraphics{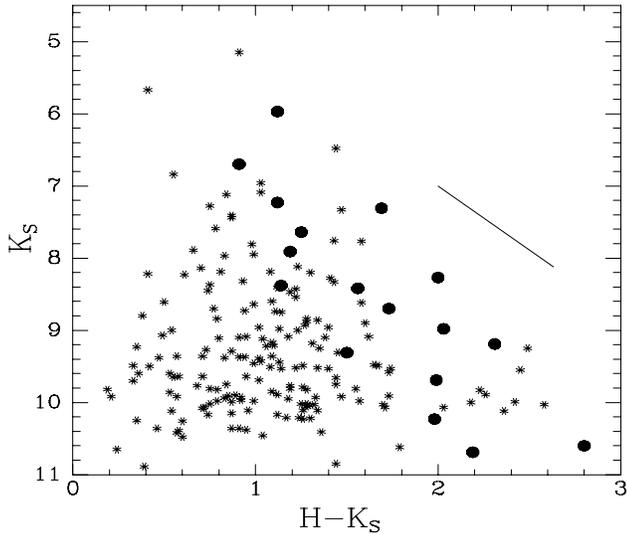}}
   \caption[]{Color-magnitude diagram for the stars in our sample.
Filled circles represent stars with $I({\rm H_2O}) > 0$
(large-amplitude variables), expected to be intrinsically very
luminous. Magnitudes are from 2MASS. The straight line represents
the displacement due to an extinction of $A_V = 10$, using the
Rieke \& Lebofsky~(\cite{rieke85}) extinction law.}
  \label{colmag}
\end{figure}

  Figure~\ref{colmag} suggests that it is possible to make
a rough luminosity estimate based on the 2MASS photometry alone.
The figure plots the position of all objects in our sample in a
$(H-K_S, K_S)$ color-magnitude diagram, including AGB stars
characterized by their prominent steam bands (see
Section~\ref{H2O_CO}), which are marked as full circles. These
stars are expected to be long-period variables roughly following
the period-luminosity relationship for Miras and semiregular
variables, which has been recently reexamined by Knapp et
al.~(\cite{knapp03}) using Hipparcos parallaxes. Their absolute
magnitudes range from $M_K \simeq -6.4$ for the shortest periods
to $M_K \simeq -8.2$ at the tip of the AGB, and thus overlap with
the absolute magnitude range of cool supergiants, from which they
can be otherwise easily separated spectroscopically. Concerning
the brightest AGB stars, their absolute $M_K$ magnitudes are
expected to depend on the metallicity, primarily due to the cooler
temperatures reached at the tip of the RGB by the most metal-rich
populations as compared to others of lower metallicity, which
result on a greater $K$-band bolometric correction (Houdashelt et
al.~\cite{houdashelt00}) combined with an essentially
metallicity-independent bolometric luminosity (Sweigart et
al.~\cite{sweigart90}). A moderate extrapolation of the
metallicity-$M_K$ relationship established by Ferraro et
al.~(\cite{ferraro00}) from a sample of globular clusters to the
solar metallicities indicates an expected $M_K = -7$ for the
brightest RGB stars, thus overlapping with the absolute magnitudes
of the AGB stars with the shortest periods.

  The positions of AGB stars, whose absolute magnitudes are confined
to a range of $\sim 2$~mag, lie in a relatively well defined band
of the $(H-K_S, K_S)$ diagram whose slope results from the
combined effect of extinction and distance on the apparent $K_S$
magnitude. The fact that the scatter within this band is of the
order of the scatter in the absolute magnitudes of Miras and
semiregular variables suggests that the amount of reddening in the
$H-K_S$ color can be used as a rough distance indicator. The
intrinsic colors of Mira variables are somewhat uncertain: Glass
et al. (\cite{glass95}) finds noticeable differences in intrinsic
colors when comparing Miras in the Large Magellanic Cloud, the
solar neighbourhood, and the galactic bulge. They find intrinsic
$H-K$ colors of bulge Miras (in the SAAO system) that range
between 0.32 and 0.61, which is redder than the reddest RGB stars
(Houdashelt et al.~\cite{houdashelt00}) and supergiants (Ducati et
al.~\cite{ducati01}). Therefore, assuming the reddening in $H-K_S$
as a proxy for the distance, RGB and supergiants having the same
position in the $K_S, H-K_S$ diagram as a given AGB star should be
more distant, and therefore more luminous.

  The lower edge of the {\it locus} of our large-amplitude variables
can be approximated by the straight line $K_S = 5.87 + 2.20
(H-K_S)$, which allows us to define a reference magnitude $K_0$ as

  $$K_0 = K_S - 5.87 - 2.20 (H-K_S)  \eqno(2)$$

\noindent whose positive or negative value indicates that the star
is either below or above the lower edge of the AGB band,
respectively. Although it may be tempting to directly relate $K_0$
to $M_K$, it must be noted that the straight line is just a
convenient approximation to describe the {\it locus} of AGB stars
in our sample and does not have a sound physical foundation.
Indeed, it is easy to show that the assumption that $K_0$ is
simply the absolute magnitude offset by a constant requires the
distance modulus to be $DM = 12.93 + 0.42 E(H-K_S)$ (where
$E(H-K_S)$ is the amount of reddening, the coefficient $0.42$
preceding it is derived from the Rieke \&
Lebofsky~(\cite{rieke85}) extinction law, and the constant after
the equal sign follows from the assumption that the lower edge of
the AGB band is defined by stars having $M_K = -6.4$), which does
not allow for distances smaller than 3.8~kpc. Still, the use of
Eq.~(2) allows the split of our sample between "bright" and
"faint" stars taking AGB stars as an appropriate dividing line.

  The fact that the bulk of stars in our sample falls below the
$K_0 = 0$ line in Figure~\ref{colmag} confirms that our sample is
dominated by giants. However, there are some bright stars with
different amounts of reddening occupying the AGB locus but with no
or weak steam absorption; note in particular the group of faint,
very red stars at $(H-K_S) > 2.0$. These are some of the most
promising M supergiant candidates in our sample.

\begin{figure}
   \resizebox{8.5cm}{!}{\includegraphics{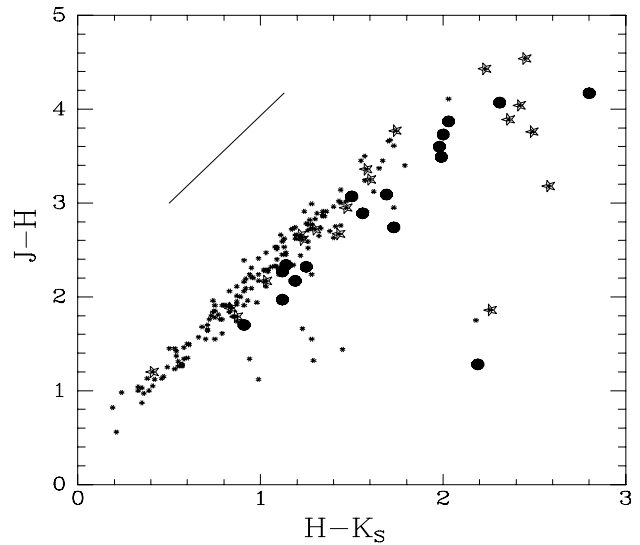}}
   \caption[]{Color-color diagram for the stars in our sample,
based on 2MASS magnitudes. Filled circles represent stars with
$I({\rm H_2O}) > 0$ (large-amplitude variables). Five-pointed
stars represent our best supergiant candidates (see
Section~\ref{cand_supergiants}). Note that the $J$ magnitudes of
stars with very red $(H-K_S)$ colors are often unreliable due to
the likelihood of contamination by nearby stars that are brighter
in $J$. The straight line represents the displacement due to an
extinction of $A_V = 10$, using the Rieke \&
Lebofsky~(\cite{rieke85}) extinction law.}
  \label{colcol}
\end{figure}

  The vast majority of the stars in our sample cluster along the
narrow strip of the $(H-K_s)$, $(J-H)$ color-color diagram that
traces the reddening vector (see Figure~\ref{colcol}). Some stars
deviate from this strip apparently occupying the region of
bluer-than-normal $(J-H)$ colors at a given $(H-K_S)$ that would
indicate the existence of near infrared excess. These are however
stars with very red colors whose faint $J$ magnitudes are near or
below the confusion limit for 2MASS in the galactic plane, for
which the 2MASS $J$-band measurement is likely to be significantly
contaminated by nearby stars. We thus consider the $J$ magnitudes
as generally unreliable for very red objects, thus precluding a
discussion on the infrared excesses of different classes of
objects.

\subsection{Equivalent widths and luminosities\label{eqw_K0}}

\begin{figure}
   \resizebox{8.5cm}{!}{\includegraphics{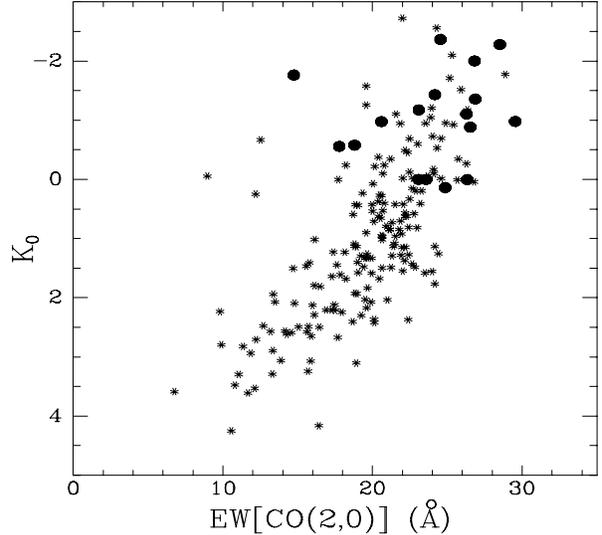}}
   \caption[]{$EW[{\rm CO(2,0)}]$ vs. the $K_0$ quantity defined
in Eq.~(2) for the stars in our sample. Filled circles indicate
stars with $I({\rm H_2O}) > 0$.}
  \label{CO20_K0}
\end{figure}

\begin{figure}
   \resizebox{8.5cm}{!}{\includegraphics{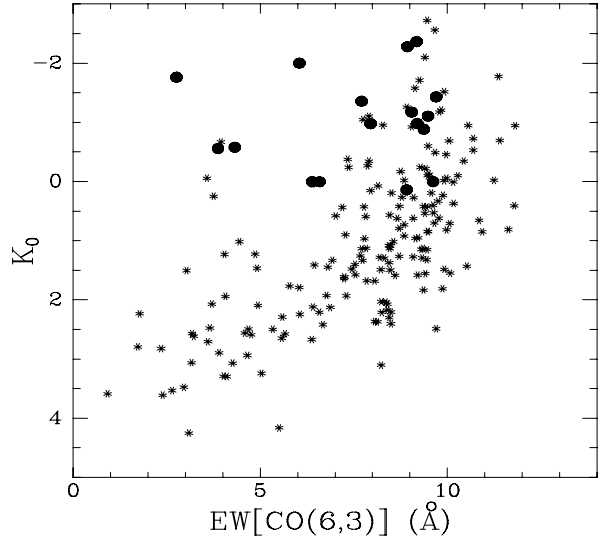}}
   \caption[]{Same as Figure~\ref{CO20_H2O}, but now with
$EW[{\rm CO(6,3)}]$ in the horizontal axis. Stars with $I({\rm
H_2O}) > 0$ tend to lie to the left due to the same veiling effect
on the CO(6,3) band that causes the diagonal band in
Figure~\ref{CO63_H2O}.}
  \label{CO63_K0}
\end{figure}

  The use of $K_0$ as defined in Eq.~(2) allows us to study
possible correlations between this quantity, purely derived from
broad-band photometry, and the equivalent widths measured on the
spectra. The most obvious correlations among both types of
quantities are shown in Figures~\ref{CO20_K0} and \ref{CO63_K0}.
While both figures are qualitatively similar in showing the
increase of average luminosity with increasing CO band depth,
there are some interesting differences between both. The CO(2,0)
band depth vs. $K_0$ diagram shows an approximately constant
slope, with considerable scatter, over the whole range of
equivalent widths covered by our spectra. Fig.~\ref{CO63_K0} shows
however a break in the trend around $EW[{\rm CO(6,3)}] \simeq
7$~\AA\, with the distribution of points flaring up to brighter
values of $K_0$. The clear displacement of the average
distribution of stars with strong steam absorption towards lower
$EW[{\rm CO(6,3)]}$ is a direct consequence of the fact, already
mentioned earlier (see also Figure~\ref{CO63_H2O}) that strong
steam absorption is correlated with weak CO(6,3) absorption,
unlike what is observed in the case of the CO(2,0) bandheads.

  Less pronounced, but still obvious correlations with $K_0$ are
found for the NaI and CaI bands. We have also looked for possible
systematic effects of ratios of equivalent widths with $K_0$ that
may provide spectroscopic discriminators, in particular the
$EW{\rm (CO(2,0))} / (EW{\rm(CaI)}+ EW{\rm(NaI))}$ ratio that
Ram\'\i rez et al.~(\cite{ramirez97}) proposed to discern  between
dwarfs and giants. The usefulness of that ratio in discriminating
also between giants and supergiants is suggested by the results of
F\"orster Schreiber~(\cite{forster00}), and has been investigated
by Schultheis et al.~(\cite{schultheis03}) for their galactic
center sample. Unfortunately we can only confirm the negative
results of Schultheis et al.~(\cite{schultheis03}), at least at
the resolution common to our spectra, in which no hints of a
systematic trend with the luminosity indicator $K_0$ is seen.

\begin{figure}
   \resizebox{8.5cm}{!}{\includegraphics{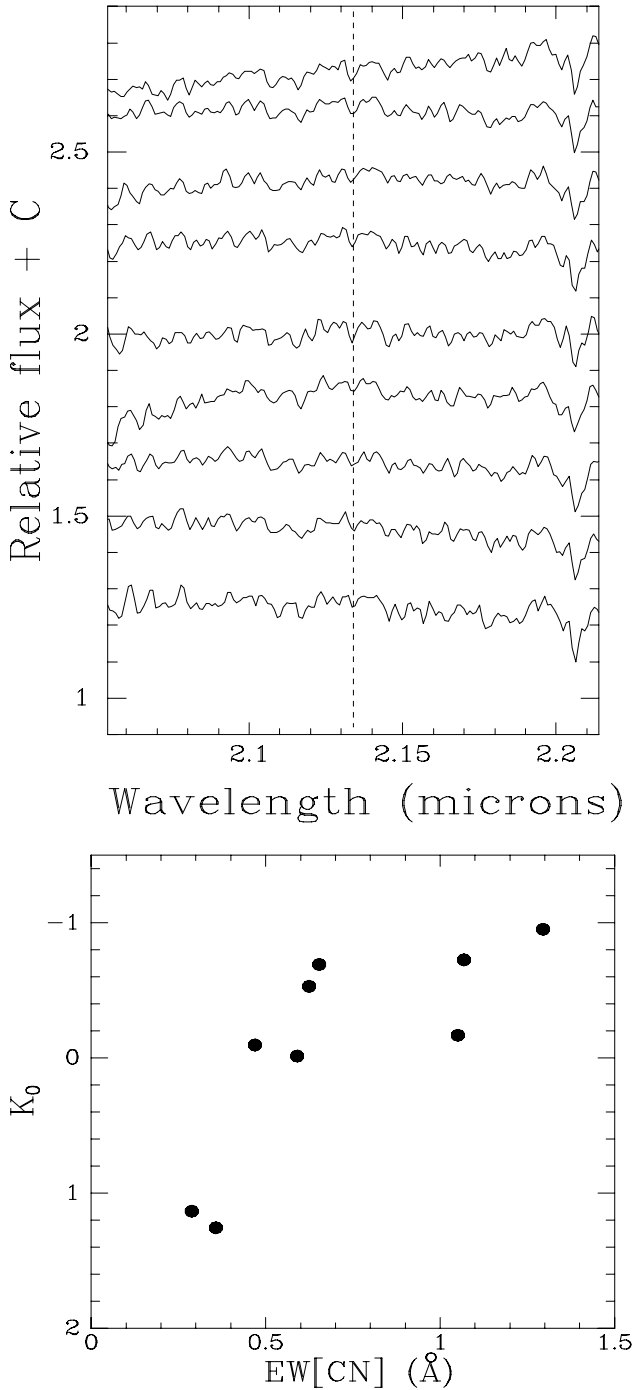}}
   \caption[]{A possible luminosity effect in the CN bands? In the
top panel, spectra are sorted by the value of $K_0$ (Eq.~(2)),
with the most negative values (implying on the average more
luminous stars) at the top. The CN feature, marked by the dashed
line, has a trend to be stronger in the most luminous spectra. The
stars plotted here are the ones in our sample for which we measure
$24 < EW[{\rm CO(2,0)}] < 25$, to remove possible temperature
effects on the CN bands. However, such trend is not confirmed for
the general sample. The bottom panel shows the equivalent width
vs. $K_0$, quantifying the effect hinted at by the top panel.}
  \label{CN_K0}
\end{figure}

  Other absorption features appearing in our spectra such as the
MgI (2.281~$\mu$m) or FeI (2.228~$\mu$m) lines may also be useful
indicators of temperature, metallicity, or surface gravity. In
particular we have attempted to use $EW[{\rm CO(2,0)]} / EW[{\rm
MgI}]$ as a possible surface gravity indicator based on the
results of F\"orster-Schreiber~(\cite{forster00}), which suggest
two different characteristic values of that ratio for giants and
supergiants at the coolest temperatures. Again we find no
systematic trend in our results, although its possible existence
may be hidden due to the weakness of the MgI feature that becomes
close to the typical $\pm 0.5$~\AA\ uncertainty in our equivalent
width measurements, thus introducing a large random scatter in the
ratios that we can determine. Similar uncertainties due to the
measurement errors relative to the small equivalent width affect
measurements involving the other lines, such as FeI (2.228~$\mu$m)
or SiI (1.590~$\mu$m)

  LW2000 have noted that the strong CN features appearing
shortwards of 1.1~$\mu$m are characteristic of the supergiants of
their sample. Unfortunately that spectral region is too heavily
obscured in the majority of our sources, thus preventing
measurements of the strongest CN bands. However, there are
multiple CN features appearing on the bluer side of the $K$ band
that may provide a useful alternative in heavily reddened objects.
We find hints of a surface gravity dependence in the blend of CN
lines near 2.134~$\mu$m, as shown in Figure~\ref{CN_K0}. Again,
the weakness of the feature prevents its accurate measurement in
many of our spectra, but higher resolution and S/N spectra might
prove this to be an efficient way of producing clean samples of
giants or supergiants. We note that the comparison of CN and CO
absorption bands provide measurements of surface abundances of
CNO-processed elements bearing important implications on the
internal structure and dredge-up history of the objects under
study (Carr et al.~\cite{carr00}; see also
Section~\ref{discussion}).

\subsection{Defining a sample of candidate
supergiants\label{cand_supergiants}}

  As discussed in the previous Sections, we have failed to find in
our spectra a discontinuity between the properties of giants and
supergiants allowing a clear discrimination between both groups.
Moreover, very few objects in our sample display simultaneously
weak or absent H$_2$O absorption wings and very strong CO(2,0)
absorption in the range covered by the supergiants of LW2000, thus
suggesting that supergiants with spectroscopic characteristics
closely matching those of the local sample are not represented in
large numbers among the stars that we observed. Nevertheless, we
do find many stars with NaI absorption similar or greater than
those of local supergiants, no signs of H$_2$O absorption wings,
moderate but not extreme CO(6,3) and CO(2,0) absorption, and
photometric characteristics that suggest an intrinsic luminosity
exceeding that of the brightest giants. These properties lead us
to suspect that our sample does contain a number of cool
supergiants in the inner Galaxy whose spectroscopic
characteristics present systematic differences with respect to
local supergiants, being in fact closer to those of the galactic
center supergiants, as noted in Section~\ref{col_mag}; see also
Section~\ref{gal_cen}. Taking together the indicators referred to
above, we have produced a list of candidate supergiants in the
inner Galaxy composed by the stars in our sample that fulfill the
following criteria:

\begin{itemize}
\item $I({\rm H_2O}) < -0.1$, denoting the absence of significant
H$_2$O absorption wings. The negative threshold is quite
restrictive, as weak steam absorption is common among the coolest
supergiants (Lan\c{c}on \& Rocca-Volmerange~\cite{lancon92}) and
the $I({\rm H_2O})$ values of the typical Mira-like spectra in our
sample reach several tenths above zero. As a matter of fact, as
many as seven of the LW2000 supergiant spectra would {\it not} be
considered as supergiant candidates on the grounds of this
criterion (see Figure~\ref{CO20_H2O}).

\item $EW{\rm [NaI]} > 4$~\AA, thus selecting as supergiant
candidates objects with a NaI feature stronger than that of the
coolest giants in metal-rich globular clusters and in the solar
neighbourhood.

\item $EW{\rm [CO(2,0)]} > 19$~\AA\ and $EW{\rm [CO(6,3)]} >
7$~\AA. We have already pointed out that these thresholds are well
below the typical strengths of the CO features in the local
supergiants, but they correspond to the values for which stars
with $K_0 < 0$ appear in Figures~\ref{CO20_K0} and \ref{CO63_K0}.
As we will discuss in Section~\ref{discussion}, the possibility
that the CO features are systematically weaker among bright cool
stars in the inner Galaxy makes it advisable to use
not-too-restrictive criteria on the CO features.

\item $K_0 < -0.1$, to select stars whose broad-band photometry
suggests high luminosity. The usefulness and limitations of $K_0$
as a luminosity indicator have been discussed in
Section~\ref{col_mag}, and we stress again here that $K_0$ is
expected to be only loosely correlated with the absolute
magnitude. We consider it however as a useful additional indicator
of the possible supergiant nature of our candidates when used in
combination with the other criteria described in this Section. The
threshold chosen is intended to exclude stars that may have been
placed above the $K_0 = 0$ line by inaccuracies in the photometry.
\end{itemize}

\begin{table*}
\caption{Candidate supergiants with no H$_2$O absorption wings}
\begin{tabular}{ccccccccccc} \hline
\noalign{\smallskip}
 & & & & & & \multicolumn{4}{c}{Equivalent widths} & \\
 Number  &      RA     &    dec      &  $K_S$ &
$J-H$ & $H-K_S$ &
CO(6,3) & NaI & CaI & CO(2,0) &$I({\rm H_2O})$ \\
 & (2000) & (2000) & \multicolumn{3}{c}{(2MASS)} & (\AA) & (\AA) &
 (\AA) & (\AA) & (mag) \\
\noalign{\smallskip} \hline \noalign{\smallskip}\\
011 & 17:59:36.64 & -23:41:16.3 &  7.33 & 2.95 & 1.47 & 11.36 &
4.38 & 4.41 & 28.87 & -0.197 \\
026 &  18:02:45.68 & -22:24:40.2 & 8.90 & 3.25 & 1.60 &  9.67 &
4.60 & 3.63 & 22.17 & -0.179\\
059 & 18:07:15.19 & -20:19:11.6 & 7.76 & 2.67 & 1.43 &  8.92 &
4.37 & 3.61 & 19.57 & -0.187\\
063 & 18:07:50.04 & -20:09:12.5 & 7.41 & 1.79 & 0.87 &  7.34 &
4.48 & 4.07 & 20.38 & -0.190\\
072 & 18:08:54.05 & -19:46:14.1 & 10.03 & 3.18 & 2.58 &  9.91 &
4.59 & 3.87 & 25.91 & -0.215\\
091 & 18:13:16.80 & -18:00:49.2 & 9.25 & 3.76 & 2.49 &  9.39 &
5.42 & 5.95 & 25.31 & -0.218\\
094 & 18:14:21.54 & -17:28:59.4 & 9.99 & 4.04 & 2.42 &  9.83 &
5.07 & 5.51 & 23.95 & -0.167\\
114 & 18:16:30.34 & -16:26:56.5 & 8.20 & 2.72 & 1.30 & 10.68 &
5.27 & 4.45 & 24.30 & -0.117\\
123 & 18:16:41.18 & -16:22:49.5 & 7.12 & 1.88 & 0.84 &  9.48 &
4.15 & 5.03 & 23.01 & -0.108\\
128 & 18:17:14.03 & -16:07:58.6 & 8.12 & 2.61 & 1.23 &  9.97 &
4.84 & 4.48 & 22.36 & -0.132\\
130 & 18:17:18.60 & -16:04:53.0 & 9.53 & 3.77 & 1.74 &  8.75 &
4.22 & 3.57 & 24.06 & -0.246\\
131 & 18:17:22.53 & -16:06:31.0 & 8.43 & 2.66 & 1.22 &  9.50 &
5.38 & 5.21 & 22.48 & -0.160\\
136 & 18:17:27.66 & -16:04:14.9 & 8.62 & 3.36 & 1.58 & 10.69 &
5.46 & 4.76 & 24.01 & -0.196\\
143 & 18:17:36.86 & -16:06:59.8 & 9.83 & 4.43 & 2.23 & 10.55 &
5.31 & 5.23 & 23.54 & -0.232\\
145 & 18:17:41.07 & -15:55:26.4 & 9.89 & 1.86 & 2.26 &  8.27 &
5.85 & 5.20 & 24.89 & -0.230\\
161 & 18:20:57.79 & -14:43:28.4 & 10.12 & 3.89 & 2.36 & 11.81 &
4.32 & 2.79 & 21.85 & -0.239\\
167 & 18:23:17.55 & -13:47:16.0 & 6.96 & 2.17 & 1.03 &  9.78 &
5.37 & 5.87 & 26.34 & -0.135\\
178 & 18:24:37.26 & -12:54:04.0 & 9.55 & 4.54 & 2.45 &  9.25 &
5.23 & 4.37 & 25.15 & -0.108\\
\noalign{\smallskip}\hline
\end{tabular}
\bigskip
\label{tab_msg}
\end{table*}

\begin{table*}
\caption{Candidate supergiants with weak H$_2$O absorption wings}
\begin{tabular}{ccccccccccc} \hline
\noalign{\smallskip}
 & & & & & & \multicolumn{4}{c}{Equivalent widths} & \\
 Number  &      RA     &    dec      &  $K_S$ &
$J-H$ & $H-K_S$ &
CO(6,3) & NaI & CaI & CO(2,0) &$I({\rm H_2O})$ \\
 & (2000) & (2000) & \multicolumn{3}{c}{(2MASS)} & (\AA) & (\AA) &
 (\AA) & (\AA) & (mag) \\
\noalign{\smallskip} \hline \noalign{\smallskip}\\
064 & 18:07:50.72 & -20:04:09.2 &  7.77 &  2.91 &  1.58 &  9.13 & 4.21 & 3.34 & 19.58 & -0.063\\
070 & 18:08:37.35 & -19:50:05.3 &  8.40 &  3.24 &  1.57 &  9.05 & 4.50 & 3.68 & 25.41 & -0.059\\
074 & 18:09:37.50 & -19:28:27.1 & 10.07 &  4.11 &  2.03 &  7.87 & 5.16 & 5.06 & 26.27 & -0.098\\
097 & 18:14:53.10 & -17:00:51.3 &  8.28 &  2.75 &  1.41 & 11.40 & 5.12 & 3.90 & 24.57 & -0.050\\
121 & 18:16:38.18 & -16:23:34.0 &  9.09 &  3.12 &  1.62 & 10.43 & 5.76 & 4.23 & 25.73 & -0.024\\
125 & 18:16:55.97 & -16:09:54.3 &  5.15 &  1.91 &  0.91 &  9.46 & 5.09 & 4.59 & 21.99 & -0.031\\
152 & 18:18:05.31 & -15:57:19.4 &  6.48 &  2.76 &  1.44 &  9.66 & 5.59 & 4.60 & 24.29 & -0.035\\
184 & 18:25:11.20 & -12:28:40.2 &  8.33 &  3.02 &  1.43 & 10.04 & 5.12 & 4.40 & 22.48 & -0.049\\
194 & 18:27:39.07 & -11:39:23.0 &  7.09 &  2.11 &  1.03 &  7.74 & 4.59 & 4.10 & 23.88 & -0.045\\
\noalign{\smallskip}\hline
\end{tabular}
\bigskip
\label{tab_msg_weakH2O}
\end{table*}

  The selection of the best cool supergiant candidates obtained
through the application of these criteria thus contains 18
objects, listed in Table~\ref{tab_msg}. We have excluded one
object, Star~191, due to its relatively blue $(H-K_S$) = 0.40; as
indicated in Section~\ref{colmag} the rough correlation between
broad-band colors and luminosity given by Eq.~(2) must break down
for blue colors indicative of low extinction, and the
characteristics of Star~191 are well compatible with those of a
lightly reddened, nearby RGB star. It is interesting to note as a
validation of these criteria that 7 out of the 8 reddest stars in
our sample, which are heavily obscured stars but still bright
enough to be in our sample and are thus good supergiant candidates
based on their photometric properties alone, do fulfill all the
spectroscopic criteria as well. Likewise, Star~011, which is the
one that best matches the spectral characteristics of the LW2000
sample of M supergiants, is also included in our selection as a
moderately reddened star and thus probably not very distant, as
may be expected from an object whose properties are similar to
those of the local sample.

  As noted above, the criterion based on the almost complete absence
of H$_2$O absorption wings is deliberately conservative, as the
coolest local supergiants do show moderate absorption. We thus
provide in Table~\ref{tab_msg_weakH2O} an additional list of M
supergiant candidates, with a lower level of confidence than those
provided in Table~\ref{tab_msg}, with weak H$_2$O absorption. Such
a table would contain most LW2000 local supergiants as well,
although their $EW{\rm [CO(2,0)]}$ would be greater on average.

  The $K_S$ magnitudes of most objects in Tables~\ref{tab_msg} and
\ref{tab_msg_weakH2O} and the reddenings implied by their $H-K_S$
are consistent with distances to the Sun placing them well within
the solar circle given the expected absolute $K$ magnitudes of
cool supergiants. However this may not be true for the faintest,
most reddened sources in those tables. For instance, assuming $M_K
= -11$ (which is very bright, although not unrealistic for a M
supergiant; Elias et al.~\cite{elias85}) and $(H-K_S)_0 = 0.3$ for
Star~072 would place it at 17.1~kpc from the galactic center on
the opposite side of the solar circle, assuming a galactocentric
distance of the Sun of 8~kpc, and obscured by $A_V \simeq 35$~mag.
Nevertheless, most of the other candidate supergiants are both
brighter at $K$ and less reddened, and the assumption that they
are indeed located in the inner galactic disk can be expected to
be generally valid. This is even more so for the RGB stars of our
sample, whose absolute magnitudes are fainter.

\section{Discussion\label{discussion}}

\subsection{Cool giants and supergiants in the galactic center
\label{gal_cen}}

  Differences similar to those that we have discussed in the
previous Sections between our sample of cool luminous stars and
local samples have been already noted in previous works on the
galactic bulge, as already discussed in Section~\ref{atomic}, as
well as on the galactic center. The latter are of particular
interest, since they include supergiants in addition to RGB stars.
Early infrared investigations of the stellar population at the
galactic center revealed a high density of cool, luminous stars
including supergiants and giants (Lebofsky et
al.~\cite{lebofsky82}; Sellgren et al.~\cite{sellgren87}). Similar
to their hotter counterparts, the comparison of properties of cool
massive stars in the galactic center with similar stars in the
solar neighbourhood has been the subject of intense study to
investigate how the widely different environment influences their
formation and evolution.

  Sellgren et al.~(\cite{sellgren87}; see also Blum et
al.~\cite{blum96}) already noted that a number of galactic center
giants and supergiants display unusually strong CaI and NaI
features when observed at low spectral resolution. Although
differences between galactic center and solar neighbourhood
luminous stars have been often interpreted as due to the effects
of metallicity on stellar evolution, detailed abundance analyses
have recently called this into question by showing that galactic
center stars have metallicities very close to solar values (Carr
et al.~\cite{carr00} and references therein; Ram\'\i rez et
al.~\cite{ramirez00a}; Blum et al.~\cite{blum03}).

  Most interestingly in the context of our results is the detailed
abundance analysis performed by Carr et al.~(\cite{carr00}) on
\object{IRS~7}, the brightest M supergiant in the galactic center.
While spectroscopy at resolution similar to ours also shows strong
CaI, NaI features for this star, its CO band at 2.293~$\mu$m is
found to be significantly weaker than those of the local M
supergiants \object{$\alpha$~Ori} and \object{VV~Cep}, whose
temperatures are very similar to that of \object{IRS~7} as shown
by model atmosphere fits. The study of Carr et al.~(\cite{carr00})
reveals a strong surface depletion of C and O and a corresponding
enhancement of N with respect to local supergiants, while the
combined abundance of C, N, and O adds to a value close to solar.
Such altered abundances not only explain the weaker CO features
(whose strength is essentially determined by the C abundance in
O-rich atmospheres), but also the unusual strength of the NaI and
CaI features in low resolution ($R \simeq 1000$) spectra. Indeed,
high resolution spectroscopy reveals that these atomic features
are strongly contaminated by a large number of CN bands, which can
even dominate the contribution to the feature in the coolest stars
(Ram\'\i rez et al.~\cite{ramirez97}). The enhancement in the
surface abundances of CNO-processed elements has the net effect of
increasing the contribution of CN to the observed NaI and CaI
features.

  The interpretation given by Carr et al.~(\cite{carr00})
invokes deep mixing enhanced by fast stellar rotation as the
explanation for the difference between galactic center and local
supergiants. They also suggest that higher rotation velocities on
the average, caused by the extreme prevailing conditions for star
formation, could also explain the differences observed in other
classes of high mass stars at the galactic center without
requiring metallicities higher than solar. A similar explanation
may also apply to the galactic center giants whose low-resolution
spectral properties are also comparable.

  As noted in Section~\ref{col_mag}, four more {\it bona fide}
supergiant candidates have been recently identified among ISOGAL
sources near the galactic center by Schultheis et
al.~(\cite{schultheis03}). Their classification as supergiants is
based on the estimated absolute magnitude and the absence of water
steam absorption. Three of these stars have $EW{\rm [CO(2,0)]}$ in
the 20-24~\AA\ interval, well within the range covered by the
supergiant candidates in our sample and also by local RGB stars.
The fourth one has $EW{\rm [CO(2,0)]} < 28$~\AA, which is also
well below those of most local supergiants in LW2000 sample. These
results thus support the trend of galactic center supergiants to
have CO bands weaker than those of local supergiants.

\subsection{Metallicity and systematic differences in feature strengths}

  It is extremely unlikely that an explanation based on rotation such
as the one proposed by Carr et al.~(\cite{carr00}) can account
also for the properties of our inner Galaxy sample, since the
formation conditions at distances of a few kiloparsecs from the
galactic center should be much closer to those in the solar
neighbourhood and no reason why our stars should be faster
rotators is apparent. However, we note that the constraint of
explaining the peculiar features of galactic center stars while
not invoking supersolar metallicity most probably does not apply
to our sample, which populates a region of the galactic disk where
the existence of a higher average metallicity is well established
by a wide variety of tracers; see Chiappini et
al.~(\cite{chiappini01}) for an exhaustive summary of galactic
abundance gradient determinations for different elements. It is in
fact very difficult to disentangle the effects of high metallicity
from those of fast rotation on stellar evolution, and in
particular on mixing in the interior of the star, as has been
discussed by Schaerer~(\cite{schaerer00}).

  The resemblance between the distinctive properties of galactic
center stars and those of our sample (greater strength of the CaI
and NaI features, comparative weakness of the $K$-band CO
absorption relative to similar stars in the solar neighbourhood)
leads us to consider it very plausible that the spectroscopic
features of many of our stars can be explained also by the
enhanced abundance of CNO-processed material in the stellar
surface resulting from deep mixing, as suggested by Carr et
al.~(\cite{carr00}) for \object{IRS~7} and possibly other galactic
center objects. However, the fact that our stars most likely have
a higher average metallicity than those of the solar neighbourhood
and also those of the galactic center leads us to consider
metallicity, rather than rotation, as the reason for the enhanced
surface CNO element abundance. Higher metallicity implies higher
opacity in the interior of the star, leading to a deeper
penetration of the outer convective zone during the first
dredge-up (Vandenberg \& Smith~\cite{vandenberg88}, Gratton et
al.~\cite{gratton00}). In turn higher mass loss rates, enhanced by
the higher metallicity, may also play a role by efficiently
removing the outer envelope and rapidly make visible at the
surface deep layers whose composition has been altered by CNO
reactions (G. Meynet, priv. comm.)

  In view of our results and the above discussion, we refrain from
deriving quantitative properties of our stars from the available
spectroscopic material using extrapolations of relationships
calibrated on populations considerably different from ours, such
as the metallicity calibration of Frogel et al.~(\cite{frogel01})
or the $EW{\rm [CO(2,0)]}$-based temperature scale proposed by
Ram\'\i rez et al.~(\cite{ramirez97}). We find it necessary to
defer a tentative determination of the intrinsic properties of our
sample (temperatures, metallicities, surface gravities, absolute
luminosities...) to future work based on high resolution
spectroscopy, which should allow quantitative analyses based on
detailed atmosphere and stellar evolution models.

\section{Summary and conclusions\label{conclusions}}

  In this paper we have presented the results of a search for
cool, luminous stars in the inner region of the Milky Way
($6^\circ < l < 21^\circ)$, selected by the signature of their
strong CO bandheads longward of $\lambda = 2.293$~$\mu$m in
narrow-band imaging. Low resolution spectroscopy has been
presented for 191 stars confirmed to have moderate or strong CO
bands and other signatures of cool spectra. Some of these stars
are readily classified as large-amplitude variable candidates from
the appearance of broad H$_2$O absorption wings. Our spectra also
allow the measurement of a number of atomic and molecular features
in the $H$ and $K$ bands. The addition of near-infrared photometry
from 2MASS, which is available for all the stars in our sample,
allows us to estimate rough luminosities. We find that our sample
is dominated by red giant branch stars, but the photometric
characteristics of some stars suggests luminosities in the range
covered by the intrinsically more luminous large-amplitude
variables, suggesting that they are red supergiants. The
spectroscopic characteristics are consistent with such
classification.

  We have compared the spectroscopic results obtained for our
sample to those derived from the atlas of near-infrared spectra of
cool luminous stars of Lan\c{c}on \& Wood~(\cite{lancon00}), which
is mostly composed of stars in the solar neighbourhood and bulge
red giants, as well as to the samples of nearby red giants of
Ram\'\i rez et al.~(\cite{ramirez97}) and bulge giants of Ram\'\i
rez et al.~(\cite{ramirez00b}). We have also made a comparison to
a sample of red giants in globular clusters with metallicities in
the $-0.37 < \rm [Fe/H] < -0.17$ from Frogel et
al.~(\cite{frogel01}). The results of this comparison may be
summarized as follows:

\begin{itemize}

\item The local sample of red supergiants displays spectra with CO
bands significantly stronger than those of the stars in the inner
Galaxy sample, especially the CO(2,0) band. The difference is less
pronounced concerning the CO(6,3) band, where there is significant
overlap between our sample and the local one. We find only one
star in our sample whose $EW{\rm [CO(2,0)]}$ is well in the range
of those of local supergiants.

\item The CaI and NaI features in the $K$ band are on the average
as strong, or even stronger, than those of local supergiants. As
compared to local and globular cluster RGB stars, most stars in
our sample have $EW{\rm [CaI]}$ and, especially, $EW{\rm [NaI]}$
well exceeding those of the coolest stars in those samples.
However, similarly strong CaI and NaI features are observed among
bulge giants. The net result of this effect and the one described
in the previous item is a well defined shift of the band defined
by our sample in the $EW{\rm [CaI]}$ vs. $EW{\rm [CO(2,0)]}$ or
the $EW{\rm [NaI]}$ vs. $EW{\rm [CO(2,0)]}$ with respect to the
average {\it loci} of the reference samples of the solar
neighbourhood.

\item An application of combined photometric and spectroscopic
criteria allows us to define a sample of 18 good M supergiant
candidates having no noticeable H$_2$O absorption, strong CO
absorption both in the $H$ and $K$ bands, and $H$, $K_S$
photometry similar to that of large amplitude variables,
suggesting that they are also intrinsically very luminous. Most of
them are considerably reddened by extinctions reaching up $A_V
\simeq 40$~mag. Nine additional candidates with weak H$_2$O
absorption wings are also identified.

\item A comparison between the spectroscopic characteristics at
low resolution of our best supergiant candidates and of the rest
of stars in our sample confirms results of other authors about the
unsuitability of spectra at $R \simeq 1000$ or lower to clearly
discern between red giants and supergiants, complicated with the
interpretation of the measurements of the CO and atomic features,
and in particular the actual species that contribute to the
latter.

\end{itemize}

  The discussion of our results (Section~\ref{discussion}) is
centered on the fact that the low-resolution characteristics noted
in our sample and summarized above (strong NaI and CaI features,
probably due to their contamination by CN lines, and comparatively
weak CO absorption) are shared by giants and supergiants in the
vicinity of the galactic center. We have referred to detailed
abundance analyses by other authors of the galactic center stars,
which indicate solar metallicities but altered surface abundances
of CNO-processed elements with respect to local supergiants,
resulting in depleted C and O and enhanced N. Systematic
differences in rotational velocities, implying deeper mixing in
galactic center stars, are invoked to account for the different
surface abundances while keeping a solar metallicity. While we
find the ultimate reason of systematic differences in rotation
velocities highly implausible to explain the characteristics of
our sample, we still consider that deeper mixing and the resulting
patterns in surface abundances are an attractive explanation. We
propose that higher metallicities in the inner regions of the
Galaxy are responsible for the deep mixing, probably as a
consequence of the increased interior opacity.

  Higher resolution spectroscopy in the infrared (the only
spectral region accessible for the study of the photospheres of
such heavily reddened objects) will be needed to perform detailed
abundance analyses as well as to possibly separate giants and
supergiants in a more reliable way than was possible with the
material currently available. Nevertheless, interesting
conclusions may already be drawn for the application of the
results of our work. On the one hand, we confirm the feature,
already suspected from studies of galactic center stars, that very
strong CO absorption longwards of 2.293~$\mu$m is a distinctive
property of local supergiants but may not apply to supergiants in
other environments. On the other hand, the usefulness of local
giants and supergiants as templates for population synthesis
studies must be regarded with some caution taking into account our
results and the ones referred to on the galactic center. Indeed,
common domains of application of population synthesis are central
starbursts and the stellar environments of AGNs, where the
conditions for star formation and evolution may be better
represented by those in the inner regions or the center of the
Milky Way. The identification of a large sample of cool, luminous
field stars such as the one presented in this paper is a necessary
first step for the understanding of their physical and chemical
properties that may emerge from the combination of future high
resolution observations and detailed modelling. Such work will
provide a database useful to improve our knowledge on the star
forming history and chemical evolution of the inner galactic disk
in the proximity of the galactic plane, as well as of a variety of
extragalactic environments where large scale massive star
formation is taking place.

\begin{acknowledgements}
  We are pleased to acknowledge the Calar Alto time allocation
committee for the generous amount of time assigned to this
project. Our special thanks to the Calar Alto staff for their
excellent support during the observations. We are also pleased to
acknowledge the support received at the La Silla observatory
during our SOFI run, especially from Ms. Karla Aubel and Dr.
Olivier Hainaut, and the warm hospitality of the ESO Guesthouse in
Santiago where much of the spectroscopic data reduction was
carried out. We are thankful to Dr. Ana E. G\'omez for her
participation in one on the Calar Alto observing runs and for her
comments on early drafts of this paper. The insightful comments of
Dr. G. Meynet on the role of the metallicity in producing the
spectral differences discussed here are greatly appreciated. We
also thank the referee, Dr. John Carr, for his constructive report
that greatly helped us to improve the contents of this paper in
many places. CC acknowledges the support of the Visiting
Scientists Program at ESO for making possible her stay in
Garching, during which part of this work was done, and financial
support from MIUR/COFIN 2003028039. JT, FF, and SR acknowledge
support from the MCYT under contract AYA2003-007736. This
publication makes use of data products from the Two Micron All Sky
Survey, which is a joint project of the University of
Massachusetts and the Infrared Processing and Analysis
Center/California Institute of Technology, funded by the National
Aeronautics and Space Administration and the National Science
Foundation.
\end{acknowledgements}


\begin{thebibliography}{}

\bibitem[2001]{alard01} Alard, C., 2001, A\&A, 379, L44.

\bibitem[2001]{alonso01} Alonso-Herrero, A., Engelbracht, C.W.,
Rieke, M.J., Rieke, G.H., Quillen, A.C., 2001, ApJ, 546, 952.

\bibitem[1989]{becker94} Becker, R.H., White, R.L., Helfand, D.,
Zoonematkermani, S., 1994, ApJS, 91, 347.

\bibitem[1989]{bessell89} Bessell, M.S., Brett, J.M., Wood, P.R.,
Scholz, M., 1989, A\&A, 213, 209.

\bibitem[1996]{blum96} Blum, R.D., Sellgren, K., DePoy, D.L.,
1996, AJ, 112, 1988.

\bibitem[2003]{blum03} Blum, R.D., Ram\'\i rez, S.V., Sellgren, K.,
Olsen, K., 2003, ApJ, 597, 323.

\bibitem[1996]{bronfman96} Bronfman, L., Nyman, L.-\AA, May, J.,
1996, A\&AS, 115, 81.

\bibitem[2000]{bronfman00} Bronfman, L., Casassus, S., May, J.,
Nyman, L.-\AA, 2000, A\&A, 358, 521.

\bibitem[2003]{bruzual03} Bruzual, G., Charlot, S., 2003, MNRAS,
344, 1000.

\bibitem[2000]{carr00} Carr, J.S., Sellgren, K., Balachandran,
S.C., 2000, ApJ, 530, 307.

\bibitem[2001]{chiappini01} Chiappini, C., Mateucci, F., Romano,
D., 2001, ApJ, 554, 1044.

\bibitem[1994]{chiar94} Chiar, J.E., Kutner, M.L., Verter, F., Leous, J.,
1994, ApJ, 431, 658.

\bibitem[1995]{codella95} Codella, C., Palumbo, G.G.C., Pareschi,
G., Scappini, F., Caselli, P., Attolini, M.R., 1995, MNRAS, 276,
57.

\bibitem[1996]{comeron96} Comer\'on, F., Torra, J., 1996, A\&A,
314, 776.

\bibitem[1996]{dallier96} Dallier, R., Boisson, C., Joly, M.,
1996, A\&AS, 116, 239.

\bibitem[1987]{dame87} Dame, T.M., Ungerechts, H., Cohen, R.S.,
de Geus, E.J., Grenier, I.A., May, J., Murphy, D.C., Nyman,
L.-\AA, Thaddeus, P., 1987, ApJ, 322, 706.

\bibitem[2001]{dame01} Dame, T.M., Hartmann, D., Thaddeus, P.,
2001, ApJ, 547, 792.

\bibitem[2001]{ducati01} Ducati, J.R., Bevilacqua, C.M., Rembold,
S.B., Ribeiro, D., 2001, ApJ, 558, 309.

\bibitem[1998]{egan98} Egan, M.P., Shipman, R.F., Price, S.D.,
Carey, S.J., Clark, F.O., Cohen, M., 1998, ApJ, 494, L199

\bibitem[1985]{elias85} Elias, J.H., Frogel, J.A., Humphreys, R.M.,
1985, ApJS, 57, 91.

\bibitem[2000]{ferraro00} Ferraro, F.R., Montegriffo, P.,
Origlia, L., Fusi Pecci, F., 2000, AJ, 119, 1282.

\bibitem[1997]{fioc97} Fioc, M., Rocca-Volmerange, B., 1997, A\&A,
326, 950.

\bibitem[2000]{forster00} F\"orster Schreiber, N.M., 2000, AJ, 120,
2089.

\bibitem[2001]{forster01} F\"orster-Schreiber, N.M., Genzel, R.,
Lutz, D., Kunze, D., Sternberg, A., 2001, ApJ, 552, 544.

\bibitem[2001]{frogel01} Frogel, J.A., Stephens, A., Ram\'\i rez, S.,
DePoy, D.L., 2001, AJ, 122, 1896.

\bibitem[1991]{green91} Green, D.A., 1991, PASP, 103, 209.

\bibitem[2001]{green01} Green, D.A., 2001, A catalogue of galactic
supernova remnants, Mullard Radio Astronomy Observatory,
Cambridge.

\bibitem[1995]{glass95} Glass, I.S., Whitelock, P.A., Catchpole, R.M.,
Feast, M.W., 1995, MNRAS, 273, 383.

\bibitem[2000]{gratton00} Gratton, R.G., Sneden, C., Carretta, E.,
Bragaglia, A., 2000, A\&A, 354, 169.

\bibitem[1992]{helfand92} Helfand, D.J., Zoonematkermani, S.,
Becker, R.H., White, R.L., 1992, ApJS, 80, 211.

\bibitem[1995]{hinkle95} Hinkle, K., Wallace, L., Livingston, W., 1995,
PASP, 107, 1042.

\bibitem[2000]{houdashelt00} Houdashelt, M.L., Bell, R.A., Sweigart,
A.V., 2000, AJ, 119, 1448.

\bibitem[1989]{hughes94} Hughes, V.A., MacLeod, G.C., 1994, ApJ,
427, 857.

\bibitem[2004]{ivanov04} Ivanov, V.D., Rieke, M.J., Engelbracht,
C.W., Alonso-Herrero, A., Rieke, G.H., Luhman, K.L., 2004, ApJS,
151, 387.

\bibitem[1986]{kleinmann86} Kleinmann, S.G., Hall, D.N.B., 1986,
ApJS, 62, 501.

\bibitem[2003]{knapp03} Knapp, G.R., Pourbaix, D., Platais, I.,
Jorissen, A., 2003, A\&A, 403, 993.

\bibitem[2003]{kolpak03} Kolpak, M.A., Jackson, J.M., Bania, T.M.,
Clemens, D.P., Dickey, J.M., 2003, ApJ, 582, 756.

\bibitem[1996]{kotilainen96} Kotilainen, J.K., Forbes, D.A.,
Moorwood, A.F.M., van der Werf, P.P., Ward, M.J., 1996, A\&A, 313,
771.

\bibitem[1992]{kuchar94} Kuchar, T.A., Bania, T.M., 1994, ApJ, 436,
117.

\bibitem[1992]{lancon92} Lan\c{c}on, A., Rocca-Volmerange, B., 1992, A\&AS,
  217, 271.

\bibitem[2000]{lancon00} Lan\c{c}on, A., Wood, P.R., 2000, A\&AS, 146, 217.

\bibitem[1989]{leahy89} Leahy, D.A., Wu, X., 1989, PASP, 101, 607.

\bibitem[1982]{lebofsky82} Lebofsky, M.J., Rieke, G.H., Tokunaga, A.T.,
1982, ApJ, 263, 736.

\bibitem[1991]{livingston91} Livingston, W., Wallace, L., 1991, An
atlas of the solar spectrum in the infrared from 1850 to
9000~cm$^{-1}$ (1.1 to 5.4~$\mu$m), NSO Technical Report, NOAO.

\bibitem[1989]{lockman89} Lockman, F.J., 1989, ApJS, 71, 469.

\bibitem[1999]{lopez99} L\'opez-Corredoira, M., Garz\'on, F., Beckman,
J.E., Mahoney, T.J., Hammersley, P.L., Calbet, X., 1999, AJ, 118,
381.

\bibitem[2002]{lopez02} L\'opez-Corredoira, M., Cabrera-Lavers,
A., Garz\'on, F., Hammersley, P.L., 2002, A\&A, 394, 883.

\bibitem[2001]{lopez01} L\'opez-Corredoira, M., Hammersley, P.L.,
Garz\'on, F., Cabrera-Lavers, A., Castro-Rodr\'\i guez, N.,
Schultheis, M., Mahoney, T., 2001, A\&A, 373, 139.

\bibitem[1996]{maiolino96} Maiolino, R., Rieke, G.H., Rieke, M.J.,
1996, AJ, 111, 537

\bibitem[1990]{mathis90} Mathis, J.S., 1990, ARA\&A, 28, 37.

\bibitem[2001]{mcclure01} McClure-Griffiths, N.M., Green, A.J.,
Dickey, J.M., Gaensler, B., Haynes, R.F., Wieringa, M.H., 2001,
ApJ, 551, 394.

\bibitem[2002]{mcquinn02} McQuinn, K.B.W., Simon, R., Law, C.J.,
Jackson, J.M., Bania, T.M., Clemens, D.P., Heyer, M.H., 2002, ApJ,
576, 274.

\bibitem[1998]{meyer98} Meyer, M.R., Edwards, S., Hinkle, K.H., Strom, S.E.,
1998, 508, 397.

\bibitem[2001]{ojha01} Ojha, D.K., 2001, MNRAS, 322, 426.

\bibitem[1996]{omont96} Omont, A., 1996, ASP Conf. Ser. 102, 305.

\bibitem[1993]{origlia93} Origlia, L., Moorwood, A.F.M., Oliva,
E., 1993, A\&A, 280, 536.

\bibitem[2000]{origlia00} Origlia, L., Oliva, E., 2000, A\&A, 357,
61.

\bibitem[2003]{picaud03} Picaud, S., Cabrera-Lavers, A., Garz\'on,
F., 2003, A\&A, 408, 141.

\bibitem[1997]{ramirez97} Ram\'\i rez, S.V., DePoy, D.L., Frogel, J.A.,
Sellgren, K., Blum, R.D., 1997, AJ, 113, 1411.

\bibitem[2000a]{ramirez00a} Ram\'\i rez, S.V., Sellgren, K., Carr, J.S.,
Balachandran, S.C., Blum, R., Terndrup, D.M., Steed, A., 2000a,
ApJ, 537, 205.

\bibitem[2000b]{ramirez00b} Ram\'\i rez, S.V., Stephens, A.W., Frogel, J.A.,
DePoy, D.L., 2000b, AJ, 120, 833.

\bibitem[1985]{rieke85} Rieke, G.H., Lebofsky, M.J., 1985, ApJ, 288, 618.

\bibitem[1997]{ruphy97} Ruphy, S., Epchtein, N., Cohen, M., Copet,
E., de Batz, B., Borsenberger, J., Fouqu\'e, P., Kimeswenger, S.,
Lacombe, F., Le Bertre, T., Rouan, D., Tiph\`ene, D., 1997, A\&A,
326, 597.

\bibitem[2000]{schaerer00} Schaerer, D., 2000, in "Stars, Gas and Dust
in Galaxies: Exploring the Links", ASP Conf. Ser. 221, 99.

\bibitem[2003]{schultheis03} Schultheis, M., Lan\c{c}on, A., Omont, A.,
Schuller, F., Ojha, D.K., 2003, A\&A, 405, 531.

\bibitem[1987]{sellgren87} Sellgren, K., Hall, D.N.B., Kleinmann, S.G.,
Scoville, N.Z., 1987, ApJ, 317, 881.

\bibitem[1987]{stetson87} Stetson, P.B., 1987, PASP, 99, 191.

\bibitem[1990]{sweigart90} Sweigart, A.V., Greggio, L., Renzini,
A., 1990, ApJ, 364, 527.

\bibitem[1998]{unavane98} Unavane, M., Gilmore, G., 1998, MNRAS,
395, 145.

\bibitem[2003]{vanloon03} van Loon, J.Th., Gilmore, G.F., Omont,
A., Blommaert, J.A.D.L., Glass, I.S., Messineo, M., Schuller, F.,
Schultheis, M., Yamamura, I., Zhao, H.S., 2003, MNRAS, 338, 857.

\bibitem[1988]{vandenberg88} Vandenberg, D.A., Smith, G.H., 1988,
PASP, 100, 314.

\bibitem[1996]{wallace96} Wallace, L, Hinkle, K., 1996, ApJS, 107,
312.

\bibitem[1997]{wallace97} Wallace, L, Hinkle, K., 1997, ApJS, 111,
445.

\bibitem[1992]{whiteoak92} Whiteoak, J.B.Z., 1992, A\&A, 262, 251.

\bibitem[1989]{wood89} Wood, D.O.S., Churchwell, E., 1989, ApJ,
340, 265.

\end{thebibliography}
\end{document}